\documentclass[sigconf,prologue,table,xcdraw]{acmart}
\usepackage{graphicx}
\usepackage[font=bf]{caption}
\usepackage{subcaption}
\usepackage{gensymb}
\usepackage{geometry}
\usepackage{enumitem}

\usepackage{amsmath}
\usepackage{booktabs} 
\usepackage{pifont}

\newcommand{\xmark}{\text{\ding{55}}}
\usepackage{xspace}
\usepackage{ctable}
\newcommand{\makeblue}[1]{\textcolor{black}{#1}}

\newcommand{\sysname}{LocAR\xspace}

\acmConference[]{}{}

\author{John Miller, Elahe Soltanaghai, Raewyn Duvall,\\ Jeff Chen, Vikram Bhat, Nuno Pereira, Anthony Rowe}
\affiliation{%
  \institution{Carnegie Mellon University}
  \streetaddress{}
  \city{}
  \state{}
  \postcode{}
}

\begin{document}

\title{Multi-User Augmented Reality with Infrastructure-free Collaborative Localization}



%
\begin{abstract}

Multi-user augmented reality (AR) could someday empower first responders with the ability to see team members around corners and through walls.  For this vision of people tracking in dynamic environments to be practical, we need a relative localization system that is nearly instantly available across wide-areas without any existing infrastructure or manual setup. In this paper, we present \sysname, an infrastructure-free 6-degrees-of-freedom (6DoF) localization system for AR applications that uses motion estimates and range measurements between users to establish an accurate relative coordinate system. We show that not only is it possible to perform collaborative localization without infrastructure or global coordinates, but that our approach provides nearly the same level of accuracy as fixed infrastructure approaches for AR teaming applications.  \sysname uses visual-inertial odometry (VIO) in conjunction with ultra-wideband (UWB) ranging radios to estimate the relative position of each device in an ad-hoc manner.  The system leverages a collaborative 6DoF particle filtering formulation that operates on sporadic messages exchanged between nearby users.  Unlike map or landmark sharing approaches, this allows for collaborative AR sessions even if users do not overlap the same spaces. \sysname consists of an open-source UWB firmware and reference mobile phone application that can display the location of team members in real-time using mobile AR.  We evaluate \sysname across multiple buildings under a wide-variety of conditions including a contiguous 30,000 square foot region spanning multiple floors and find that it achieves median geometric error in 3D of less than 1 meter between five users freely walking across 3 floors.

\end{abstract}

%
%

%

%
\maketitle

\section{Introduction} \label{sec:introduction}

Driven by advances in visual-inertial odometry (VIO), simultaneous localization and mapping (SLAM) and miniaturized depth sensing technologies, we are seeing augmented reality (AR) technologies becoming more accessible on a wide variety of platforms.  Mobile phones are being equipped with specific dedicated hardware to enable richer AR experiences, including multiple cameras, specialized processors, UWB ranging radios \cite{u1} and small LIDAR depth sensors.  Navigation applications like Google Maps AR mode, Ikea Place, and games like Pokemon Go have shown some of the early potential of AR on mobile phones.

While these single-user applications have largely been successful, developing interactive multi-user applications has proven to be substantially more difficult.  Multi-user AR presents a unique set of challenges involving communication, synchronization, and localization between users. Overcoming such challenges opens the door for truly groundbreaking applications to emerge, both in mobile AR and for wearable headsets.  Take for example a first responder or firefighter application, where teams of users navigate through a previously unexplored or harsh (damaged/modified) environment.  With the availability of a robust multi-user AR platform, it would be possible to annotate people and their paths, and drop virtual markers in the environment in an entirely infrastructure-free manner.  First responders could use headset AR to see the status and position of fellow teammates and the location of support vehicles even through walls without any {\em a priori} scene information. In the mobile phone context, this same type of platform could even help you find a friend at a concert venue or your keys at home.

\makeblue{In order to overlay virtual content that appears from the user’s perspective to be "anchored" to the physical world, it is necessary to track the pose of the user’s display relative to the world.  As the user rotates/translates the display, the projected content needs to move accordingly, which requires accurate 6-degree-of-freedom (6DoF) motion tracking.  With a single user, it is sufficient to perform this tracking with respect to any arbitrary starting pose.  The position and orientation of the origin is irrelevant, as long as the current pose is accurate with respect to that origin.  With multiple users, the problem becomes more challenging – in order for each user to see the same virtual content at the same physical location, each tracking instance must share the same 6DoF origin.  This requires some collaboration between the devices. While gravity provides a reference direction for one axis (provided the devices are equipped with accelerometers), magnetic field readings are inconsistent in indoor environments and thus unreliable as a yaw reference \cite{rajagopal2019improving}.}

\makeblue{Current AR frameworks, such as Apple’s ARKit and Google’s ARCore, as well as off-the-shelf headsets like Microsoft’s Hololens 2, now have provisions to enable multi-user applications~\cite{swiftshot}.  While each uses a slightly different approach, all rely on sharing visual (and depth) features between users in order to establish a common coordinate system.  As each user detects distinguishable features in the environment, these features are collected into a map using SLAM.  By sharing this map, other users are able to localize themselves if they detect the same visual features.  While we are optimistic about the future of multi-user vision-based SLAM, the current frameworks currently fall short in terms of reliability and scalability across large areas.  Finding a visual feature match that is robust enough to provide a common origin currently requires the two users to view the surrounding scene from a very similar perspective.  However, in many scenarios, visual matching between the users is not possible even in close proximity. For example, if the users view the scene from different directions or under different lighting conditions, or if objects in the scene have been moved or become occluded, visual feature matching struggles. Additionally, in search and rescue applications, users are often purposefully taking disjoint paths through the environment to improve coverage, so there will be no common visual features for map matching (either because they are separated by walls or are in a visual denied environment with smoke). In order to provide building-scale coverage and beyond, it is necessary to maintain a large and (ideally) dense feature map, which quickly becomes impractical to store and share. } 

\begin{figure}
    \centering
    \includegraphics[width=\columnwidth]{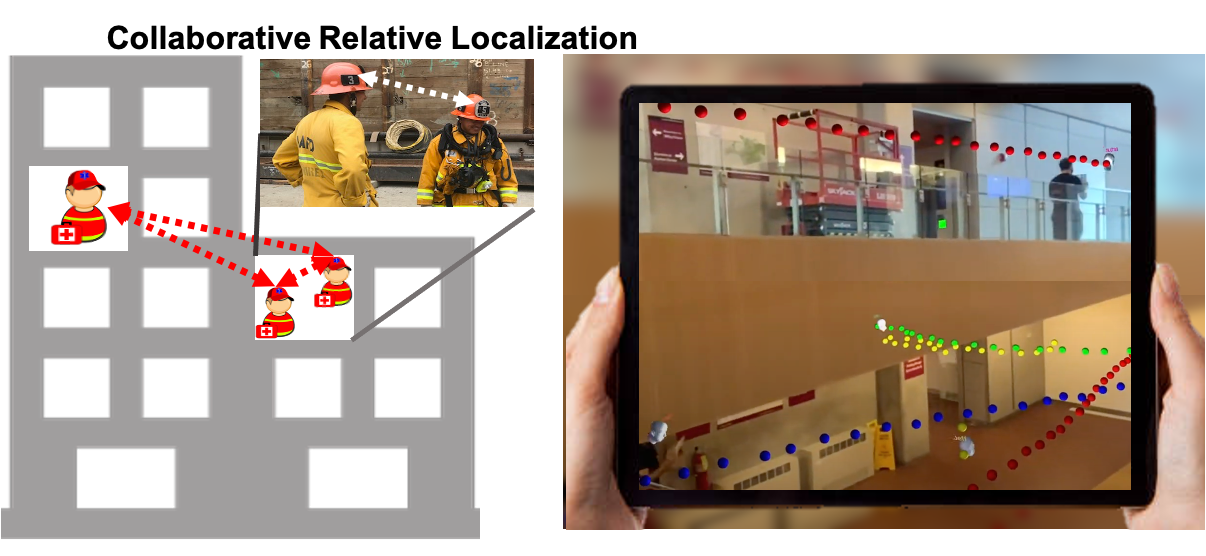}
    \caption{\sysname offers a distributed infrastructure-free relative localization framework that allows multiple mobile users to create a collaborative AR session.}
    \label{fig:intro}
\end{figure}

This paper addresses these limitations by proposing \sysname, a distributed relative localization framework that allows multiple AR users to create an on-demand collaborative AR session without any prior infrastructure. \sysname uses motion information from VIO and ranging measurements between users from UWB radios, which are now available on the latest generation of mobile phones and specialized AR headsets, to estimate the relative pose (6DoF) of each AR user. \sysname's key innovation is the design of a collaborative particle filter that jointly estimates the 6DoF pose of all AR users relative to each other without requiring any map sharing or pre-existing localization infrastructure. Since it does not rely on sharing visual features, this approach is broadly applicable to static or dynamic environments, both indoor and outdoor, including search-and-rescue scenarios where the need for visual pre-mapping is a nonstarter. \sysname is an alternative technique for setting up multi-user AR sessions by sharing inertial data and ranges as opposed to sharing nearby landmarks. 

To achieve this, \sysname captures the local inertial information from each individual AR user, providing the 6DoF pose of this user over time. While tracking motion, \sysname collects distance ranges (using UWB) to other users, and combines these information sources using a particle filter. Like most inertial tracking systems, VIO tracking estimates are smooth and locally accurate, but drift over time and provide no initial pose estimate. UWB ranges are infrequent and noisy, but provide absolute distance information that does not drift over time. By combining these complementary sensors, we achieve the best of both worlds. The absolute nature of UWB ranges allows us to correct VIO drift over time, while noise of UWB readings is smoothed by the VIO. In addition, the distributed architecture of \sysname allows each user to locally estimates the pose of other AR users with minimal message exchange between users in favor of scalability.

\makeblue{One core challenge, however, is the state-space explosion of the particle filter as the number of AR users grows, since they all must be tracked simultaneously. To address this challenge, \sysname takes a collaborative particle filtering approach that still tracks all nodes jointly, but uses Rao-Blackwell factorization to reduce the number of particles required to a tractable level. Compared to a more traditional particle filter where each user is independent, we show that the collaborative approach is able to leverage the synergistic information present between ranges to different nodes to improve accuracy, while maintaining a reasonable memory footprint that grows linearly with the number of tracked nodes. In addition, our filter formulation allows for sporadic UWB and VIO updates, loosening communication constraints in the system design over methods that rely on fixed-rate updates.}

In order to demonstrate relative user tracking and a prototype teaming use-case, we developed a mobile AR application for iOS. Our technique is fundamental to any relative tracking system that has inertial data along with ranging estimates and hence could be applied to AR headset in hands-free applications like aiding first responders (firefighters would not use mobile phones inside burning buildings). Since UWB APIs are currently not available to mobile phone developers, we created a peer-to-peer ranging firmware for the MDEK1001 evaluation module from Decawave.  The firmware allows a phone to pair with the MDEK module over BLE which in turn discovers and ranges with any number of nearby UWB devices.  Each module can be paired with mobile phones or powered by batteries to act as a stand-alone tag or beacon. The firmware is able to multiplex a BLE connection with the phone while simultaneously performing low-power neighborhood discovery using a scalable rate-adaptive round-robin protocol for ranging (discussed in more detail in Section \ref{sec:ranging_protype}).  

\makeblue{We evaluated the performance of our system in a number of environments and in four different buildings.  We tested in static as well as more dynamic environments with moving people, furniture and changing lighting.  One of our test included 5 users moving around a large (30,000+ sq ft) contiguous 3-floors area within an office building. The test environment includes long corridors, different sizes of rooms separated by concrete, drywall and various other construction materials. In a number of tests, we moved furniture and toggled lighting to simulate more dynamic elements often found in the wild.  In each test, the users walked freely creating many NLoS scenarios with multiple walls between users. Across all these experiments, \sysname provides a mean 3D geometric error performance of 0.9 m between users across 12 different random walking traces, creating over 200 groundtruth-ed points and walking periods between 5-20 minutes per test. In addition, we observe that the quality of AR performance is sensitive to more than just geometric localization error. Camera lens parameters, bearing and distance combine to define visual registration errors that are highly dependant on the scene geometry.  To better capture these effects, we also evaluate our system in terms of pixel error, which more accurately captures the visual displacement errors experienced by users. We observe that \sysname provides significantly lower pixel error compared to baseline methods that only rely on visual features. Our application source and UWB firmware is all open-source and will be available on GitHub.}


\vspace*{0.05in}\noindent\textbf{Contributions:} Our core technical contributions are:
\begin{itemize}
    \item A distributed Rao-Blackwellized Particle Filter (RBPF) formulation and implementation for real-time 6DoF relative localization.
    \item An energy-efficient peer-to-peer UWB protocol with open-source firmware tailored toward wide-area relative localization.
    \item An end-to-end implementation and thorough evaluation of an infrastructure-free AR localization system that can support multiple users. A short demo of the system in real-time can be seen here: \url{https://www.youtube.com/watch?v=5vjKUjgLqhc}. 
\end{itemize}

\section{Related Work} \label{sec:background}

\begin{table}[t]
\centering
\begin{center}
\begin{small}
	
\begin{tabular}{p{2.5cm}|p{1.5cm}|p{2cm}|p{1.5cm}}
\toprule 
Approach & \hfil \footnotesize{Infrastructure-free} &  Robustness (light, motion, etc) & \footnotesize{Computational Complexity}\\

\hline
 
 \footnotesize{Beacons / Markers \cite{wang2016apriltag,zhao2020relative}} & 
\hfil  \cellcolor[HTML]{FFFFFF}{\color[HTML]{FE0000} \xmark} &  
\hfil  \cellcolor[HTML]{FFFFFF}{\color[HTML]{FE0000} \xmark} &  
\hfil  \cellcolor[HTML]{FFFFFF}{\color[HTML]{009901} \checkmark}\\ \hline
 
\footnotesize{SLAM w beacons \cite{song2019uwb}} &
\hfil \cellcolor[HTML]{FFFFFF}{\color[HTML]{FE0000} \xmark}& 
\hfil \cellcolor[HTML]{FFFFFF}{\color[HTML]{009901} \checkmark} &
\hfil \cellcolor[HTML]{FFFFFF}{\color[HTML]{009901} \checkmark} \\ \hline

\footnotesize{Dead Reckoning\footnotemark \cite{kendall2015posenet}} & 
\hfil  \cellcolor[HTML]{FFFFFF}{\color[HTML]{009901} \checkmark}  &
\hfil \cellcolor[HTML]{FFFFFF}{\color[HTML]{FE0000} \xmark} &
\hfil \cellcolor[HTML]{FFFFFF}{\color[HTML]{009901} \checkmark} \\ \hline

\footnotesize{SLAM map sharing \cite{arkit}} & 
\hfil \cellcolor[HTML]{FFFFFF}{\color[HTML]{009901} \checkmark} &
\hfil \cellcolor[HTML]{FFFFFF}{\color[HTML]{FE0000} \xmark} &  
\hfil\cellcolor[HTML]{FFFFFF}{\color[HTML]{FE0000} \xmark}  \\ \hline

\textbf{\sysname} & 
\hfil \cellcolor[HTML]{FFFFFF}{\color[HTML]{009901} \checkmark} & \hfil \cellcolor[HTML]{FFFFFF}{\color[HTML]{009901} \checkmark}&
\hfil \cellcolor[HTML]{FFFFFF}{\color[HTML]{009901} \checkmark}
\end{tabular}

\end{small}
\end{center}
\captionof{table}{Existing localization techniques vs.\ \sysname}
\label{tab:table-comparison} 
\end{table}

\footnotetext{requires known start point for localization}

The topic of indoor localization has seen many solutions over the past decade for mobile phones, in robotics, and more recently for AR. These methods can be broadly categorized under \emph{vision-based} solutions that mainly rely on cameras, LIDARs, along with supporting sensors such as IMUs, and \emph{range-based} solutions that use RF, acoustic, infrared, or UWB beacons for ranging and location estimates. In the following sections, we discuss the related work in each of these categories and explain their limitations for multi-user AR applications.



\subsection{Vision-based Localization}
Fiducial markers such as ARTags \cite{mulloni2011handheld,klopschitz2007automatic} and AprilTags \cite{wang2016apriltag,kallwies2020determining,zhao2020relative} are frequently used in AR systems to provide a reference between the physical environment and virtual objects. While these passive markers can be accurately localized with only a camera and low computational requirements, they are not suitable for real-time tracking of AR users, especially in mobile and NLoS scenarios. 

Simultaneous localization and mapping (SLAM) techniques are widely used in robotics and AR systems for identifying and then leveraging features in an environment to track the position of a moving device. These methods use either monocular cameras \cite{bae2016fast,engel2014lsd,mur2015orb,aider2005model}, depth cameras \cite{bae2014rapid,lu2014image,yuan2016rgb}, or stereo cameras \cite{bellavia2013robust,brand2014stereo, you1999hybrid} to extract visual features from the scene and extract the 3D coordinates of the features and the device 6DoF pose. These coordinates, however, are only relative to an origin point, typically the start point. More recently, ARKit \cite{arkit} by Apple and ARCore \cite{arcoreanchors} by Google have shown persistent AR by providing 6DoF pose estimation with respect to a previously acquired map by combining vision or point-clouds with VIO. However, the biggest challenge facing vision-based localization is that it relies heavily on recognizing known images or clusters of unique feature points in the environment. This results in slow convergence and a high sensitivity to environment dynamics such as displacement of furniture, ambient lighting, and requires rich visual (or depth map) features. Even with advanced hardware platforms such as the Hololens 2 that uses depth sensors, users must often walk around and view several areas of a scene before localization is able to take effect. Due to the limitations of purely vision-based approaches, we advocate combining visual localization approaches with range-based beacons, and mainly focus on relative location of the users with respect to each other.

\makeblue{
\subsection{{Range-based Localization}} 
 Beacon-based solutions provide continuous localization using  UWB~\cite{olsson2014cooperative}, BLE~\cite{shao2018marble,cheung2006inexpensive}, or ultrasound~\cite{gomez2013indoor,lazik2015alps,li2018automatic} ranging. However, all of these methods rely on pre-installed infrastructure and dense deployments throughout the building, which is not suitable for all AR applications. An alternative approach is to combine vision and ranging mechanisms, where ranging information from on-board radios such as UWB, Bluetooth, or WiFi is used to eliminate the accumulated errors of odometry sensors~\cite{song2019uwb,olsson2014cooperative,liu2017cooperative,dhekne2019trackio,gentner2017simultaneous,wang2017ultra,rajagopal2019improving}. While probably the best systems in terms of performance, the existing solutions still require pre-installed infrastructure or known starting points to link local maps to the physical space. \sysname addresses these limitations by combining relative ranging of the users with local motion information of each users, thus achieving the best of both worlds. One recent example of using VIO with UWB for ranging was introduced by the Apple AirTag platform.  AirTag leverages a single moving phone along with UWB ranging (from the Apple U1 chipset) to detect a small battery-powered tag.  The system currently only operates in 2D and hence does not work across multiple floors and does not support a network of moving users each localizing each other.  }




\subsection{{Multi-User Localization}} 
 While typical SLAM-based solutions assume a single user, recent developments in AR frameworks, such as Google’s ARCore / Cloud Anchors, Apple’s ARKit and Microsoft's Spatial Anchors, have enabled multi-user capabilities. In these systems, each AR device individually performs SLAM to capture the visual features of the physical space relative to its local coordinate system. The users then share these visual maps to establish a common coordinate system and estimate the pose of other users. To share these maps between the users, Google ARCore uses a cloud-based architecture, which combines these maps centrally and sends the updated maps to all the users. However, Apple ARKit uses a peer-to-peer architecture, where the host of the AR session shares its current map with the users joining the session. However, any of these techniques impose significant communication overhead \cite{ran2019sharear}. These maps consist of dense visual features, 3D meshes, or raw point clouds, which are usually large and difficult to transfer. In addition, these map matching approaches assume a significant overlap between all of the users, which becomes unwieldy in terms of network traffic and computation in large areas.  On the other hand, in search and rescue environments, where multi-user AR can bring significant added value, users are often purposefully taking disjoint paths through the environment to improve coverage, thus making map matching much more challenging and substantially increasing the convergence time.

\subsection{{Relative Localization}} \label{sec:collaborative}
Traditional localization systems typically have the goal of estimating the "absolute" location in a fixed coordinate system that is mapped to the physical space using external systems. 
In this sense, the idea of "localization" is inherently tied to the existence of some form of infrastructure from which to base the measurements. However, reliance on infrastructure is infeasible in many AR scenarios, especially in the presence of multiple users. In these cases, it is instead possible to determine the \textit{relative} locations between users to establish a common coordinate system for multi-user AR applications. For example, to display a virtual overlay of a target on the screen, only the knowledge of the target's position relative to the display system is required. 




The concept of relative localization is first used in sensor network localization for collectively locating stationary \cite{nagpal2003organizing,savvides2001dynamic} and mobile \cite{moore2004robust,rad2011cooperative,eren2004rigidity} nodes with respect to each other. These works provide the theoretical foundation for network localization using graph theory \cite{eren2004rigidity,popescu2012tracking,ferner2008cooperative,vcapkun2002gps,jamali2012cooperative,barooah2007estimation}, information theory \cite{savic2013cooperative,ross2014mutual,charrow2014approximate,hoffmann2006mutual}, or Bayesian inference methods \cite{nilsson2013recursive,carlone2011simultaneous}. However, the majority of these systems are only evaluated in simulation and do not provide the desired AR performance in terms of latency (real-time operation), accuracy, and degrees of rigid body movement in space (6DoF). 

Relative localization has also been explored in robotics for localizing a swarm of drones or multiple robots with respect to each other \cite{coppola2018board,carlone2011simultaneous,olsson2014cooperative,liu2017cooperative}. However, all of these systems assume short ranges with all the nodes in Line-of-Site (LOS) and only focus on 2D or 3D localization, instead of 6DoF. In addition, most of these methods require the prior knowledge of the initial position, which suffers from accumulated error over time. Therefore, they can not be directly extended to wide-area AR applications. In this paper, we present a distributed relative localization framework that provides the real-time relative pose estimates of AR users without requiring any pre-existing infrastructure, prior mapping, or known initial position.

\vspace*{0.05in}
\noindent\textbf{Relative Localization Architectures}
In terms of information sharing, two types of relative estimation architecture can be employed for a relative localization system -- centralized and distributed. In a centralized architecture \cite{nagpal2003organizing}, all the nodes in a network collect and combine the information in a central server for fusion. This requires all nodes to be in constant communication with the central node, which results in a large communication overhead. A distributed architecture \cite{moore2004robust} doesn't require a central server; instead, each node processes the information locally using their on-board solver. The main advantage is that it is more scalable to larger network sizes at the expense of a slight reduction in accuracy.




\section{System Overview} \label{sec:system}


In this paper, we propose a distributed relative localization framework that allows multiple users to create a collaborative and persistent AR session in a completely infrastructure-free setup. The system is computationally practical to a large number of users and wide-area environments. We now describe our formulation of the localization problem in greater detail and introduce the components and algorithms we use in the system, with details being discussed in greater depth in later sections.



\begin{figure}
    \centering
    \includegraphics[width=\columnwidth]{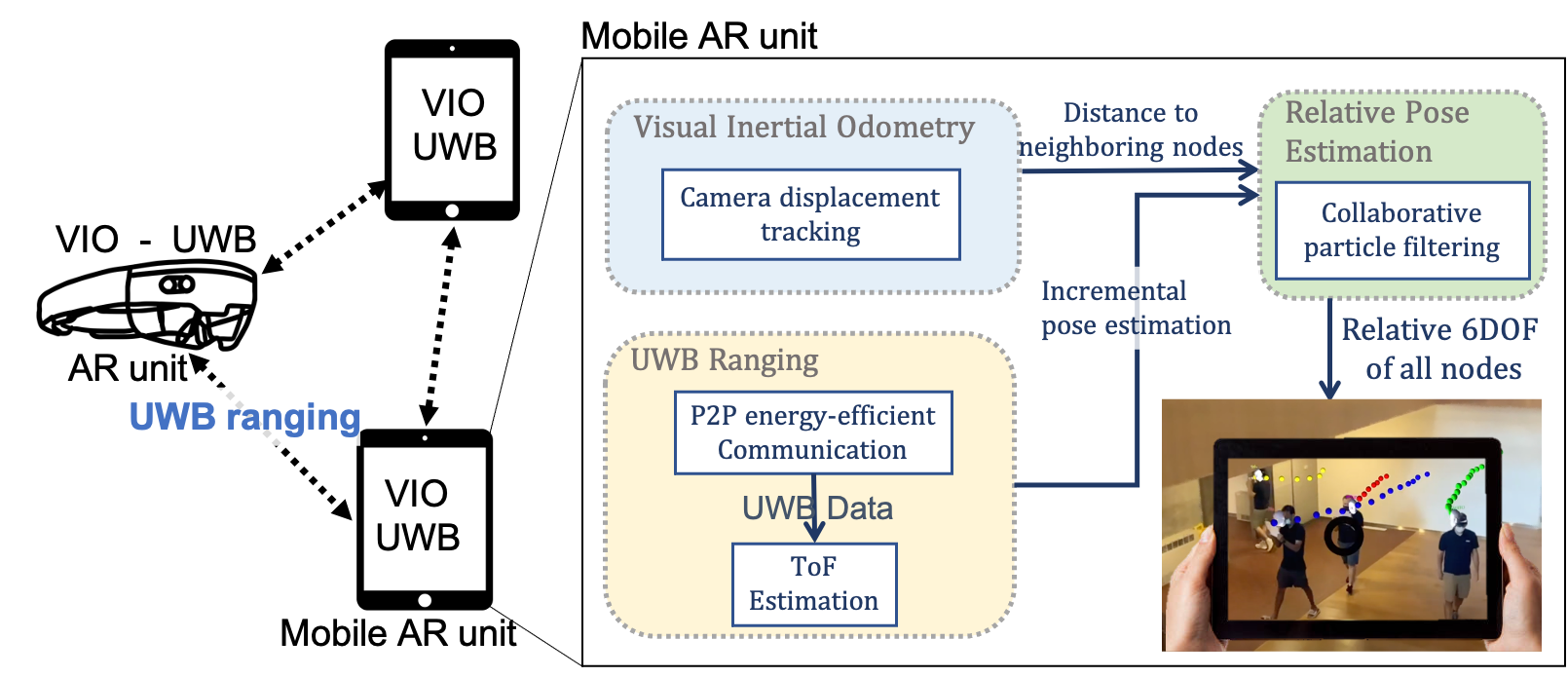}
    \caption{System Overview. A mobile user's device locates several other target devices using a combination of VIO tracking and UWB ranging. 6DOF relative tracking enables AR overlays to be drawn on the display.}
    \label{fig:overview}
\end{figure}

\subsection{Problem Formulation} \label{sec:problem}
We consider an indoor scenario consisting of $N$ mobile users (nodes), all with unknown positions and orientations. Each mobile user has an AR display device, defined as \emph{display device}, and wants to localize all other users, defined as \emph{target devices}, with respect to itself, without requiring any {\em a priori} knowledge of the physical space or pre-installed infrastructure. The localization framework has to work in real-time on limited compute platforms and needs to scale feasibly with number of devices being tracked. All AR devices are equipped with 2 sensors:
\begin{itemize}[noitemsep,topsep=0pt,leftmargin=*]
    \item \noindent\textbf{VIO tracking:} currently many smart phones and most AR headsets in the market have built-in VIO. VIO essentially tracks the motion of camera by fusing detected visual feature points with inertial sensor data. The output of VIO is the position and orientation of the device with respect to the reference frame at startup by performing conventional dead-reckoning. Even though VIO provides the camera displacement over time, there is no common origin between multiple users to extract their relative positions. Another challenge of VIO data is the accumulated drift error over time and sensitivity of visual dead-reckoning to environment conditions, such as lighting and motion. 

    \item \noindent\textbf{UWB ranging:} Among various wireless technologies that can go through obstacles such as Bluetooth, UWB, WiFi, etc, UWB is the most promising technology to combat multipath propagation in cluttered environments\cite{rajagopal2019improving}. As a result, we are seeing appearance of UWB chips on the latest mobile phones, providing peer-to-peer ranging. However, each UWB node is only capable of measuring the distance to neighboring nodes that are in range. Given the mobility of users, we cannot assume that range measurements occur synchronously or with any sort of regularity, resulting in sparse and inadequate data for real-time localization. 
    
    \makeblue{
    \item \noindent\textbf{Data Communication:} We assume that each user's device can communicate their state information with any neighbors in a peer-to-peer manner.  This requires relatively low data rate exchanges and could leverage the UWB transmissions or use an ad-hoc method like WiFi Direct, Bluetooth or dedicated emergency responder radios. One of the key benefits of our collaborative filtering approach is that devices only needs to exchange data with their neighbors that are replying to UWB messages (not a fully connected network). In our experimental platform, each node communicates using WiFi or LTE from the mobile device, but this could be easily replaced in a production implementation.  In our experiments, the state data transmitted on each active neighbor link between neighbors was below 16 kbit/s (assuming 10Hz updates).  Our system is also resilient to message drops and reasonable levels of jitter (tens of ms).  With message latencies on the order of 100ms, our system appears to perform well and is within common bounds for most single-hop wireless communication systems. 
}

    
\end{itemize}




\subsection{System Architecture} \label{sec:architecture}
We demonstrate that the synergy between VIO and UWB data allows us to overcome the challenges and limitations of each. This results in a distributed relative localization framework that allows multiple mobile users to create a collaborative AR session. First, \sysname adopts a \emph{distributed architecture} in favor of scalability by providing each node with peer-to-peer distance measurements to other nodes. Next, to deal with the sparsity of UWB reading, range measurements are combined with local camera VIO traces. The absolute nature of UWB ranging allows us to correct VIO drift over time. Finally, \sysname leverages the presence of multiple users and their mobility to collaboratively estimate the relative position of all users, improving the overall localization accuracy while maintaining low computational overhead. An overview of \sysname framework is depicted in Figure \ref{fig:overview}. Upon startup of an AR app, the AR session tracks the pose of this device using VIO from the AR API and collects the UWB ranges from neighboring nodes, which are then passed to a particle filter state-estimation solver to extract the location and orientation of other nodes with respect to itself. The next section elaborates on \sysname's collaborative pose estimation technique and the underlying challenges.

   



\subsection{Overlaying Virtual Objects} \label{sec:overlaying}

To display a virtual object, three matrices are required:
\begin{itemize}
    \item $K$: The 3x4 matrix encoding the intrinsic properties of the virtual camera such as resolution, focal length, and centerpoint, which which are assumed to be known
	\item $D_O$: The 4x4 matrix encoding the 6DOF pose of the display relative to some arbitrary origin
	\item $V_O$: The 4x1 vector encoding the 3DOF position of the target object relative to that same origin.
\end{itemize}

From these, we can calculate the pixel coordinates of the virtual object on the display $[u, v]^T$ as:
\begin{equation}
	[u',v',w']^T = K * D_O^{-1} * V_O
\end{equation}
\begin{equation}
	[u, v]^T = [u', v']^T / w'
\end{equation}
We can simply combine the latter two matrices as:
\begin{equation}
	V_D = D_O^{-1} * V_O,
\label{eq:relErr}
\end{equation}
where $V_D$ is now the 4x1 vector representing the position of the target object \textit{relative} to the display. There is no longer a requirement for a fixed origin. The coordinate system is typically chosen with +x pointing towards the right edge of the display, +y pointing towards the top edge of the display, and +z pointing out of the display towards the viewers eyes.

\section{Relative Pose Estimation}
\label{sec:PF}
Here we describe our localization framework, which uses a particle filter for tracking the $N-1$ target devices $V_D^{(i)}$ relative to the display device. First, we start by explaining a simple approach that tracks each $V_D^{(i)}$ independently and then demonstrate how it can be enhanced by tracking all $V_D^{(i)}$ jointly, using Rao-Blackwellized particle filtering (RBPF) to ensure that the problem remains tractable as $N$ grows.

\subsection{Particle Filter (PF) Formulation} \label{sec:pf_formulation}
We begin by describing the state-space representation and our error models for VIO and UWB measurements, for a basic particle filter formulation. A particle filter for our state estimation has the following benefits: (i) it is computationally easy to run online, (ii) it allows us to use arbitrary noise models to describe VIO and UWB errors, (iii) it can work with as few as 1-2 beacons in under-defined cases, (iv) as it is agnostic to update rate, it allows handling of asynchronous ranges from the beacons and does not require receiving synchronized ranges to perform trilateration.

\subsubsection{State-Space} \label{state-space}
We wish to track each device $V_D^{(i)}$ relative to the display. Each $V_D$ consists of three positional components, $x$, $y$, and $z$. In addition, since the VIO estimates from each device are with reference to a separate origin with a separate orientation, we need to add components to the state-space to track the orientation of each device as well. However, since VIO provides an orientation estimate that is gravity-aligned (thanks to the inclusion of accelerometer measurements), we need only track a single angle $\theta$ about the vertical axis for each $V_D$, where $\theta$ is the angular offset between the target device's origin orientation and the display device's. Thus, our state-space for each tracked device has 4 dimensions: $x^{(i)}$, $y^{(i)}$, $z^{(i)}$, and $\theta^{(i)}$.

\subsubsection{VIO Measurements} \label{vio}
VIO, like other forms of odometry, tracks a device's motion over time relative to some arbitrary origin. It measures $dx$, $dy$, and $dz$. Although AR frameworks on mobile devices normally perform loop closure to help mitigate drift, there is still a steady accumulation of integration error that occurs in practice, both in position and orientation about the vertical axis \cite{rajagopal2019improving}. We model these errors as Gaussian with small standard deviations $\sigma_{xyz}$ and $\sigma_\theta$, respectively. The state update equations for VIO at time $t$ are:
\begin{equation}
    x^{(i)}(t+1) = x^{(i)}(t) + dx * \cos\theta^{(i)} + dz * \sin\theta^{(i)} + N(0, \sigma_{xyz}^2)
\end{equation}
\begin{equation}
    y^{(i)}(t+1) = y^{(i)}(t) + dy + N(0, \sigma_{xyz}^2)
\end{equation}
\begin{equation}
    z^{(i)}(t+1) = z^{(i)}(t) + dz * \cos\theta^{(i)} - dx * \sin\theta^{(i)} + N(0, \sigma_{xyz}^2)
\end{equation}
\begin{equation}
    \theta^{(i)}(t+1) = \theta^{(i)}(t) + N(0, \sigma_\theta^2)
\end{equation}
We note that, although the linear velocity error $\sigma_{xyz}$ and rotational velocity error $\sigma_\theta$ are modeled as Gaussian in our formulation, any unexpected error in VIO (such as large jumps) can be recovered from using resampling techniques that account for the possibility of these jumps (see "kidnapped robot problem" in  \cite{thrun2002probabilistic}).

\subsubsection{UWB Measurements}
UWB measurements occur frequently but sporadically between pairs of nodes. They give a measurement of the distance between a pair of nodes, with an error that is roughly Gaussian with standard deviation $\sigma_r$. In the particle filter, we use a uniform model for UWB range error that extends $\pm3\sigma_r$, and assume there is a $P_{nlos}$ chance that the UWB range is entirely wrong due to non-LOS (NLOS) errors. The probability model for obtaining a UWB range $z$ to node $V_D^{(i)}$ is:
\begin{equation}
  P(z) =
  \begin{cases}
                                   P_{nlos} & \text{if $|\|V_D^{(i)} - D\| - z| > 3\sigma_r$} \\
                                   1 - P_{nlos} & \text{otherwise}
  \end{cases}
\end{equation}
where $D$ is the position of the display device relative to its starting point, as measured by its own VIO tracking.

The reason for using a uniform probability model instead of a Gaussian is to account for the conditional dependence between consecutive measurements between the same pair of nodes. Since the particle filter assumes consecutive measurements are independent, it will interpret repeated measurements as "new" information, averaging their errors together. If a Gaussian error model is used, these repeated measurements will lead to false convergence and particle impoverishment, which can be avoided by using a uniform model.

\begin{figure}
    \centering
    \includegraphics[width=\columnwidth]{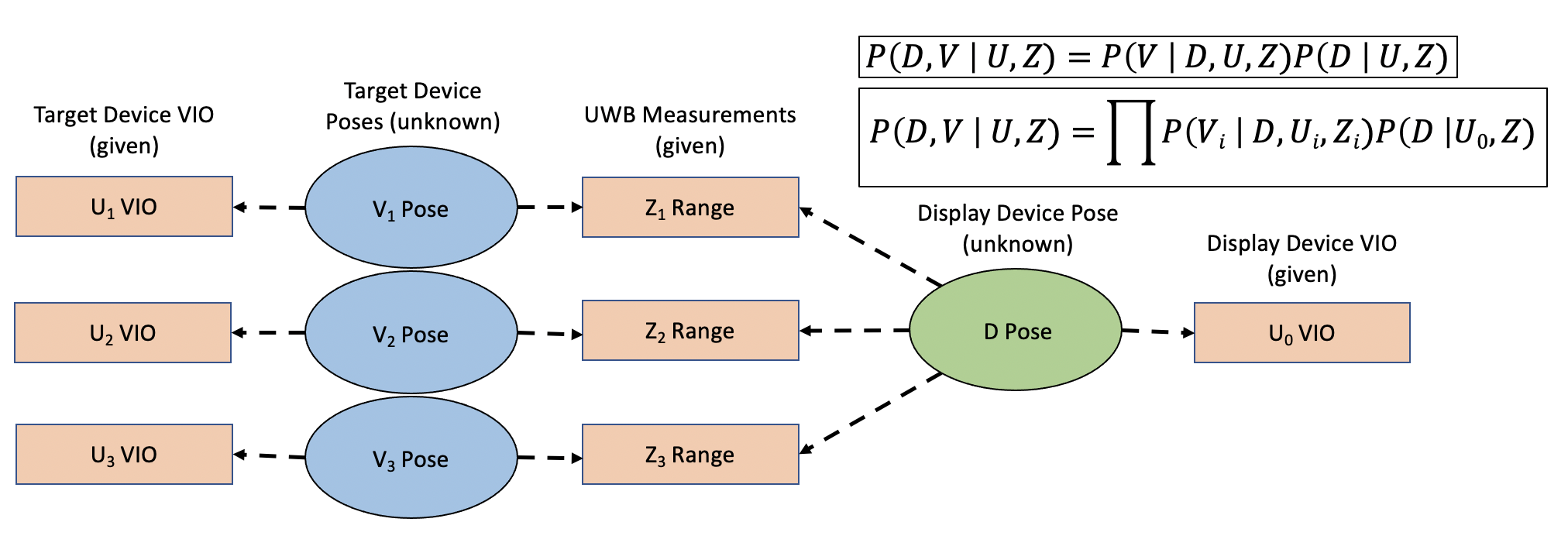}
    \caption{\sysname's Particle filter formulation jointly estimates multiple user positions ($D$ and $V_i$) by combining UWB ($Z_i$) and VIO ($U_i$) measurements. The measurement dependency graph illustrates that each $V_i$ is conditionally independent given $D$, since each UWB measurement $Z_i$ depends only on the pose of $V_i$ and $D$ (no UWB measurements are taken between $V_i$ and $V_j$). Using Bayes' rule, the joint distribution can be factorized as shown, resulting in the Rao-Blackwellized formulation in the box.}
    \label{fig:particlefilter}
\end{figure}

\begin{figure*}
    \centering
    \includegraphics[width=.8\linewidth]{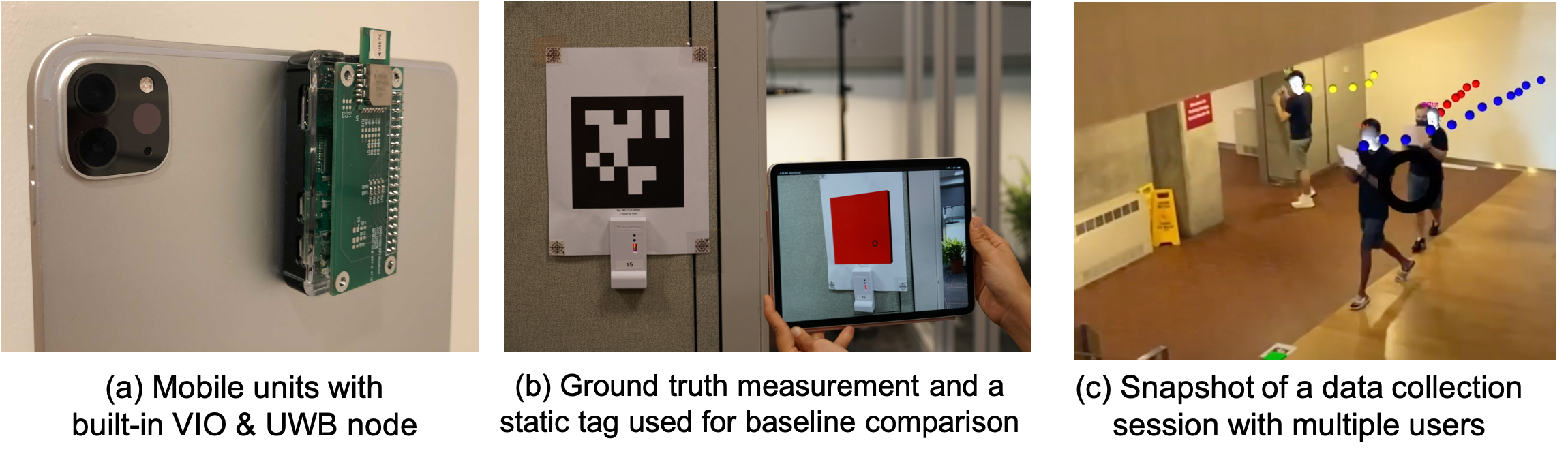}
    \caption{System Components Including UWB ranging nodes, Ground-Truth Data Collection Markers, and Mobile Tablet Application}
    \label{fig:implementation}
\end{figure*}

\subsection{Collaborative Estimation with RBPF} \label{sec:collab_est_rbpf}
In our naive baseline implementation, we run a completely independent particle filter for each device we wish to track. As a result, the computational load scales linearly with the number of devices $N$. However, since the particle filters are run independently, any error that is accumulated due to noise in the display device's {\em own} VIO tracking cannot be mitigated through collaboratively ranging to multiple devices and leveraging the synergistic information that arises.

With unlimited computational resources, it would be possible to {\em jointly} model the states of all $N$ moving devices. This way, every range could be used to improve the state estimation of all nodes in the joint distribution. However, sampling from $4N$ dimensional state-space would require a number of samples exponential in $N$ in order to adequately sample the growing dimensionality.

A solution to this problem arises when some state variables $Y^{(i)}$ are always conditionally independent given some other state variable $X$. When this is the case, it is possible to factorize the joint probability distribution and independently track each $Y^{(i)} | X$. This approach, called Rao-Blackwellization (RBPF), is common in the SLAM literature \cite{thrun2002probabilistic} as a means of estimating a map whose elements are conditionally independent given a user's location. As illustrated in Figure \ref{fig:particlefilter}, our formulation of the relative localization problem fits this framework, since UWB provides measurements of device locations that are conditionally independent given the location of the display device.

In the RBPF formulation, a particle filter is used to represent the belief of the display device $D$, where each target device $V_D^{(i)}$ can be represented by any probabilistic distribution. We chose to also represent the target device estimates using particle filters. In Section \ref{sec:collab_impact}, we demonstrate the benefit of the collaborative nature of our joint RBPF formulation over the more common naive independent particle filter approach.









\section{System Implementation} \label{sec:system_impl}

There are three main components to the implementation of our system: UWB ranging platform, AR application, and large-scale ground-truth collection. The UWB ranging platform allows collection of range data between users in a dynamically sized ad-hoc network. The AR application overlays a digital objects on the estimated relative locations in the field of view of the display, allowing users such as firefighters to know where their team is without having a direct visual. It also collects visual data from VIO and  ground-truth by decoding AprilTags, which are  placed strategically around the building to determine error in our system during data collection.

\begin{figure*}
    \centering
    \includegraphics[width=.9\linewidth]{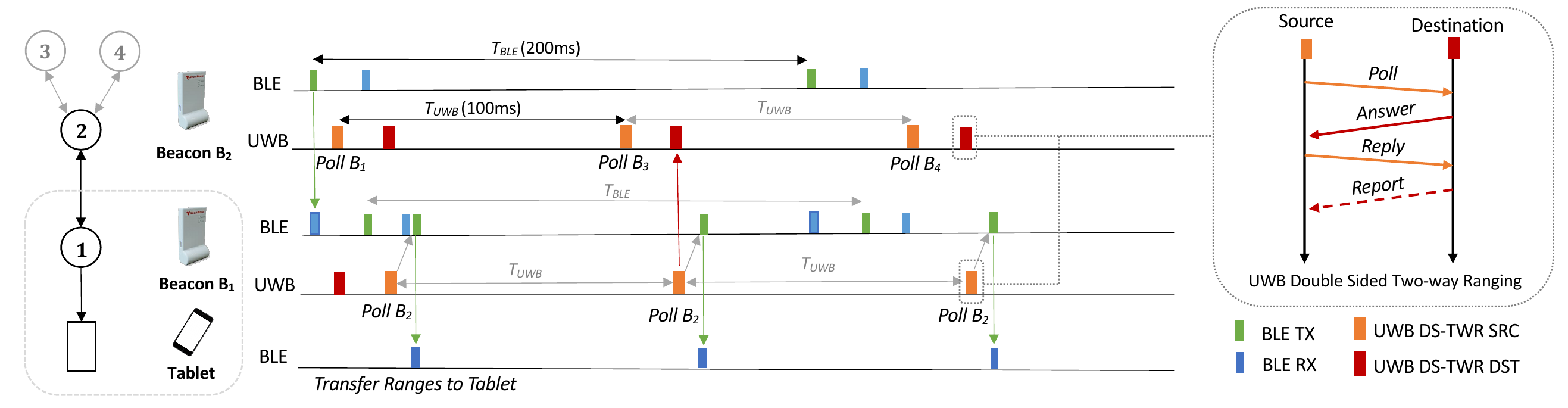}
    \caption{\sysname's BLE Neighbor Discovery and UWB Ranging Protocol allows energy-efficient peer-to-peer measurements while minimizing networking collision. }
    \label{fig:beluga-mac}
\end{figure*}

\subsection{UWB Ranging Prototype} \label{sec:ranging_protype}

Though not the focus of this paper, we realized that many localization researchers have struggled to find a UWB solution that can easily operate in peer-to-peer mode at scale. Unfortunately, most of the freely available reference implementations are designed for fixed infrastructure scenarios that support mobile devices, as opposed to fully peer-to-peer operation.  We imagine that phones with UWB hardware could eventually implement this functionality on-board once APIs become available. 

In order to perform neighborhood discovery and ad-hoc ranging with UWB, we developed an open-source and easy-to-use firmware image for the MDEK1001 modules from DecaWave. The MDEK1001 is an all in 1 battery- or USB-powered module with an enclosure that pairs a Nordic nRF52832 MCU with a DW1000 chip.  The Nordic chip has a 64 MHz Arm Cortex-M4 processor with integrated BLE radio that can be programmed to act like a stand-alone beacon or pair with a mobile phone. Our firmware image exposes a standard serial interface (the AT command set) with the ability to store default parameters to flash memory, making it easy to configure addresses, sleep modes, neighborhood discovery polling rates, and UWB ranging options. Our firmware provides three basic functionalities: (1) Neighborhood discovery using BLE's GAP discovery protocol, (2) coordinated Double-sided Two-way Ranging (DS-TWR) using UWB and (3) an interface to external systems using either USB serial or a standard BLE GATT server.  We design our protocol under the assumption that we have a highly dynamic mesh of nodes with hidden terminals and asymmetric links that change on the order of seconds. 

The neighborhood discovery protocol is BLE's standard device discovery protocol that frequency hops across three channels. We allow users to define a custom advertisement period $T_{BLE}$ (default period of 200 ms) and a configurable signal strength (RSSI) threshold for determining the most recent and closest neighbors.  To save power when nodes are idle, we duty-cycle background scanning and disable the UWB radio.  A node in the system can announce that it wants to participate in active ranging through its BLE advertisements. This in turn will wake-up all nearby nodes and activate their UWB radios. Figure \ref{fig:beluga-mac} shows an overview of the BLE and UWB transactions required to perform neighborhood discovery and ranging. Note that the BLE discovery uses three channels and not just a single channel.  Once activated, each node initiates a DS-TWR request (detailed in the upper right of the figure and in this application note \cite{APS013} ) over UWB at a user-configurable timing interval $T_{UWB}$, with a default value of 100 ms.  Each $T_{UWB}$ period the node performs a new DS-TWR request to the next node in its local neighbor list.  

If DS-TWR messages are dropped either due to collision or packet corruption, the next polling interval is randomly offset to avoid repeated collisions.  We use an exponential random distribution across $T_{UWB}$ similar in nature to Slotted ALOHA.  As one would expect, as the number of neighbors increases, the polling rate of each individual neighbor decreases.  We provide users with a lookup table for $T_{UWB}$ values needed to support particular maximum node densities within a single collision domain. In Figure \ref{fig:beluga-mac}, you can see that Beacon $B_1$ transmits every $T_{UWB}$ to Beacon $B_2$ since it has no other neighbors.  Beacon $B_2$ cycles through 3 total neighbors in its neighborhood list (the neighbor graph shown on the left). After nodes stop transmitting active ranging advertisements for a defined timeout, nodes return to their lower powered duty-cycled listening state.  As shown in the bottom line of Figure \ref{fig:beluga-mac}, we also support simultaneously pairing an actively scanning node with a mobile device using a standard BLE GATT server.  It is also possible to connect the MDEK1001 to a host device over USB serial or through its built-in RPI header.  The default parameters of our firmware support 16-bit addressing (over 30K nodes) with cluster densities of 10 nodes at approximately 1 Hz update rates for each neighbor.  Our low power sleep energy is on the order of 10 mW (mostly consumed by background BLE scanning) with an average active ranging energy of 800 mW.  In practice, we see BLE neighbor discovery on the order of a 1-2 s with a typical 10-20 s eviction timeout.  All source and documentation are available on GitHub.

\subsection{Prototype AR Application} \label{sec:proto_ar}
We developed a prototype of \sysname as a mobile AR application running on iOS. This application provides two main features: (1) it shows the relative location and orientation of other users in the scene in AR (shown in Figure \ref{fig:implementation}-c), and (2) it coordinated ground-truth data collection among mobile users (shown in Figure \ref{fig:implementation}-b).  The mobile app collected VIO data using Apple's ARKit and UWB ranging data using a MDEK1001 module from DecaWave over BLE. All ranging and communication information was shared using MQTT over WiFi, but this could conceptually be replaced by WiFi Direct or some similar peer-to-peer protocol.  ARKit captures VIO data at 60 Hz and we collected UWB ranges with a polling rate of 10 Hz.  \makeblue{ While it is difficult to exactly isolate how much energy VIO consumes on iOS and Android, the Intel T265 stands at a good reference consuming less than 1.5W.  This is low enough that it does not significantly impact interactive usage over a few hours during an AR session. }  As described above, the actual rate at which UWB data was received by each mobile user depends on the distance and number of neighbors around a particular node.


\subsection{Ground-Truth Collection}
\label{sec:GT}
One of the biggest challenges for assessing the performance of a 6DOF localization system at scale is accurately collecting ground-truth poses. We developed a data collection framework that periodically guides users to converge on "check-in" locations where AprilTags could be used to accurately record pose. We first installed over a dozen 8.5 by 11 in AprilTags~\cite{wang_16} across the multiple floors of our test building with retro-reflective markers on each corner. We surveyed the corners of each AprilTag using a total station with an estimated accuracy on the order of millimeters. To coordinate synchronized ground-truth readings between different users, we integrated AprilTag decoder into the AR application, in which the users are instructed to move to the nearest AprilTag and wait until all users across the building had a high-confidence ground-truth measurement. Given the known tag location and the pose estimated by the AprilTag decoder~\cite{wang_16}, the application computes the ground-truth location which is then published over MQTT to a central logging service.

\section{Evaluation}
\label{sec:evaluation}
\makeblue{In this section, we first explain our experimental setup, which is performed across several environments, and define our evaluation metrics. We then present the performance of \sysname and analyze the sensitivity of our system under various real-world conditions. Then, in Section \ref{sec:sensitivity}, we describe additional tests that we performed to evaluate the sensitivity of the system to different environments (including changes in lighting and background motion in the scene) as well as other factors such as user walking patterns and RF line-of-sight conditions.}

\subsection{Experimental Setup}
\label{sec:exp_setup}

\begin{figure}
    \centering
    \includegraphics[width=\linewidth]{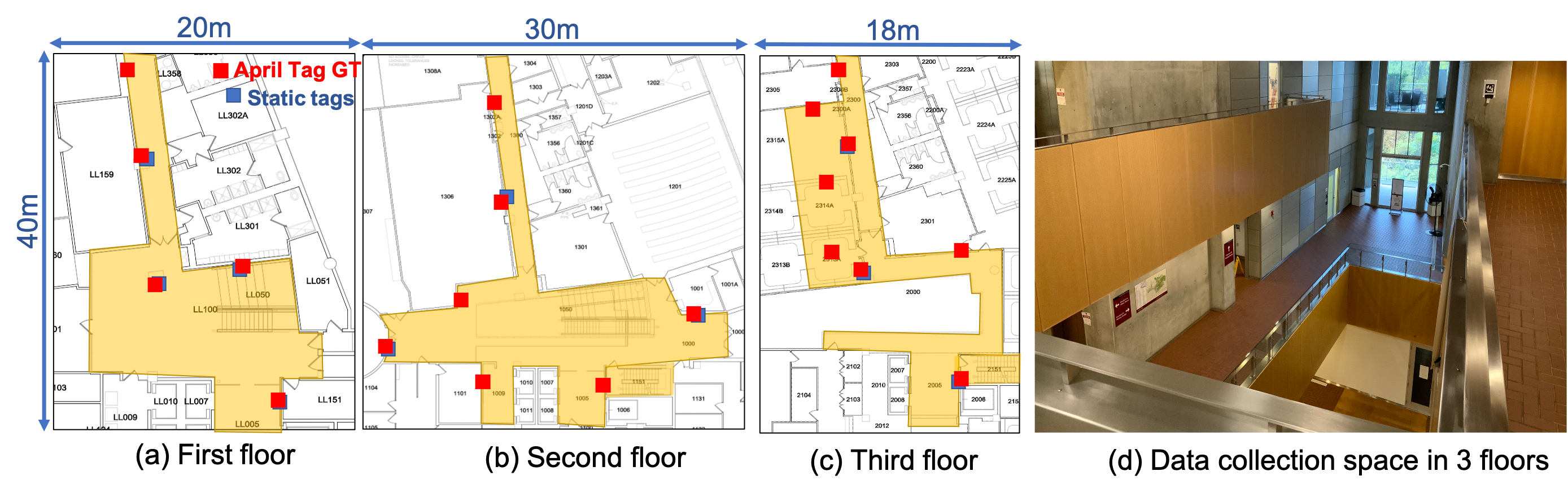}
    \caption{\makeblue{Example experimental setup for large contiguous multi-floor office environment}}
    \label{fig:exp_loc}
\end{figure}

\begin{figure}
    \centering
    \includegraphics[width=\linewidth]{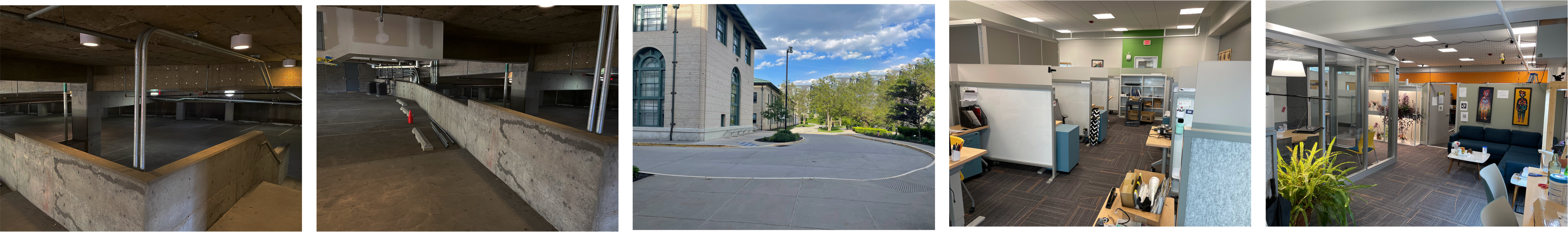}
    \caption{\makeblue{Snapshots of tested environments with different lighting and multipath conditions. }}
    \label{fig:moreEnv}
\end{figure}

\makeblue{Our primary evaluation of \sysname consisted of a deployment across a 30,000 sq ft area spanning 3 floors of an office building, with 3 to 5 users walking in an arbitrary fashion, and 9 static tags deployed for baseline comparison, as shown in Figure \ref{fig:exp_loc}. We also stress tested \sysname in a diverse set of environments both indoor and outdoor, different lighting conditions, as well as dynamic environments. The snapshots of these environments are shown in Figure \ref{fig:moreEnv}.} In all these experiments, each user carries an iPad or iPhone with a built-in VIO tracking, and a UWB node attached to the back of the device (as shown in Figure \ref{fig:implementation}-a), while the static tags consist of just the UWB platform. As noted before, \sysname does not require any pre-installed infrastructure or static beacons with known location for localization, and here the static tags are only used for our baseline comparison. Unless otherwise specified, all of our presented results only use ranges from mobile nodes.


The experiments consist of both LOS and heavy NLOS situations, with many instances where users are spread across 3 different floors with one or more dry/concrete walls between them. No instructions are provided to users on how to walk or how to hold the tablets. For 7 different experiments and 10-15 min per run, the users walk with different speeds and periodically stand stationary, resulting in a total of about 40 min worth of data per person. This data is divided into an "evaluation" set, where users are walking normally, and a "sensitivity analysis" set, where users are walking in pre-defined patterns (evaluated in Section \ref{sec:sensitivity}). As explained in Section \ref{sec:GT}, the ground-truth was obtained with a number of AprilTags surveyed in a global coordinate frame using a total station. To synchronize the ground-truth measurements between users, the AR application guides users to scan a nearby AprilTag every 5-30 s over the course of each experiment.

\subsection{Evaluation Metrics}
\label{sec:metric}
\begin{figure}
    \centering
    \includegraphics[width=.8\linewidth]{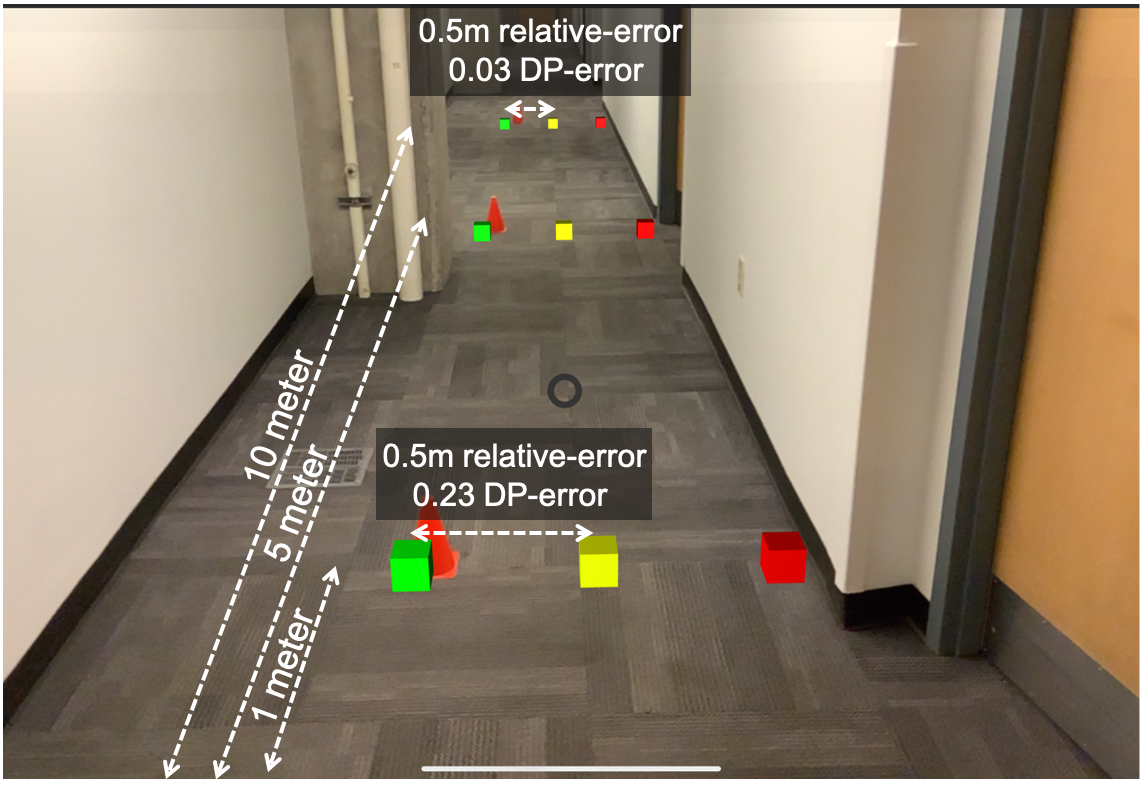}
    \caption{Virtual objects overlaid with identical geometric errors that yield dramatically different visual offsets in AR captured by display-proportional (DP) error.}
    \label{fig:ARmetric}
\end{figure}

The quality of AR performance is sensitive to more than just geometric error. Camera lens parameters, bearing, and distance combine to create the visual error seen by a user. To better capture these effects, we introduce a new AR-specific metric, which we call display-proportional error (DPE), that combines distance, bearing, and the camera field-of-view as a single cohesive benchmark. Figure \ref{fig:ARmetric} shows a typical example where 3 virtual cubes are overlaid at a fixed distance from a set of (real) physical orange cones.  The cones are located at a distance of 1, 5, and 10 m, respectively, away from the camera.  The green cube has no error, the yellow cube is offset by 0.5 m and the red cube is offset by 1 m. Notice that, due to perspective, the cubes that are further from the camera appear closer to the cone, even though their relative error in meters is the same. This simple example highlights why geometric error alone does not do justice to AR localization performance. Instead, display-proportional error computes the AR error as the distance between an object's true location and its estimated location {\em when projected onto a 2D display}, as a proportion of the display's horizontal size. In the example in Figure \ref{fig:ARmetric}, the closest yellow box has a DPE of 0.23\footnote{after accounting for the height that the user is holding the device off the ground}. This error corresponds to approximately 1/4 of the screen width, while the farthest yellow box has a DPE of only .03, or about 1/33 of the screen width. In this sense, DPE captures the reprojected error of the estimated 3D locations, and can easily be used to calculate pixel error by simply multiplying by the display's horizontal resolution.

Therefore, We formalize our error metric definitions as: 
\begin{itemize}

\item 
\noindent\textbf{3D geometric error:}
We calculate the average pair-wise Euclidean distance in 3D between all pairs of mobile nodes in meters.

    \item \noindent\textbf{Display-Proportional Error:}
\begin{equation}
    \frac{\epsilon_{xy}}{|dist + \epsilon_z|} * \frac{f_x}{H_x},
\end{equation}
Where $\epsilon_{xy}$ is the $xy$ component of the 3D geometric error, $\epsilon_z$ is the $z$ component, $dist$ is the true distance between the display and the target object, $f_x$ is the camera's focal length (in pixels), and $H_x$ is its horizontal resolution (in pixels)

\end{itemize}

\subsection{Baselines} \label{sec:baseline}
We compare the performance of \sysname with two baselines: 

\textbf{(1) VIO-Only:} a typical infrastructure-free localization method \cite{bloesch2015robust} that uses VIO for pose estimation relative to the start point.  This is a common localization method in robotics, but it requires initialization and {\em a priori} knowledge of users' start points. Even though this assumption is not feasible for most multi-user AR applications, it allows us to more easily isolate the performance contributions from VIO and UWB ranging in our system.

\textbf{(2) UWB-VIO infrastructure-based oracle:} an infrastructure-based localization technique, which uses VIO to estimate 6DOF motion and UWB ranging to fixed beacons.  We assume each fixed beacon (9 total) has a known global location in order to provide a baseline \cite{wang2017ultra}.  As shown in Figure \ref{fig:implementation}-b, static UWB nodes are placed at a fixed offset from each AprilTag to provide a set of known UWB locations. 
We consider this technique as our oracle and show that \sysname can achieve performance at nearly the same accuracy without relying on any pre-installed infrastructure.

It should be noted that both of these baselines are originally proposed for absolute localization, either relative to origin or relative to the physical coordinate system, so we obtain the relative localization for comparison with our system using Equation \ref{eq:relErr}.


\begin{figure}[t]
    \centering
     \begin{minipage}{0.49\linewidth}
        \includegraphics[width=\linewidth]{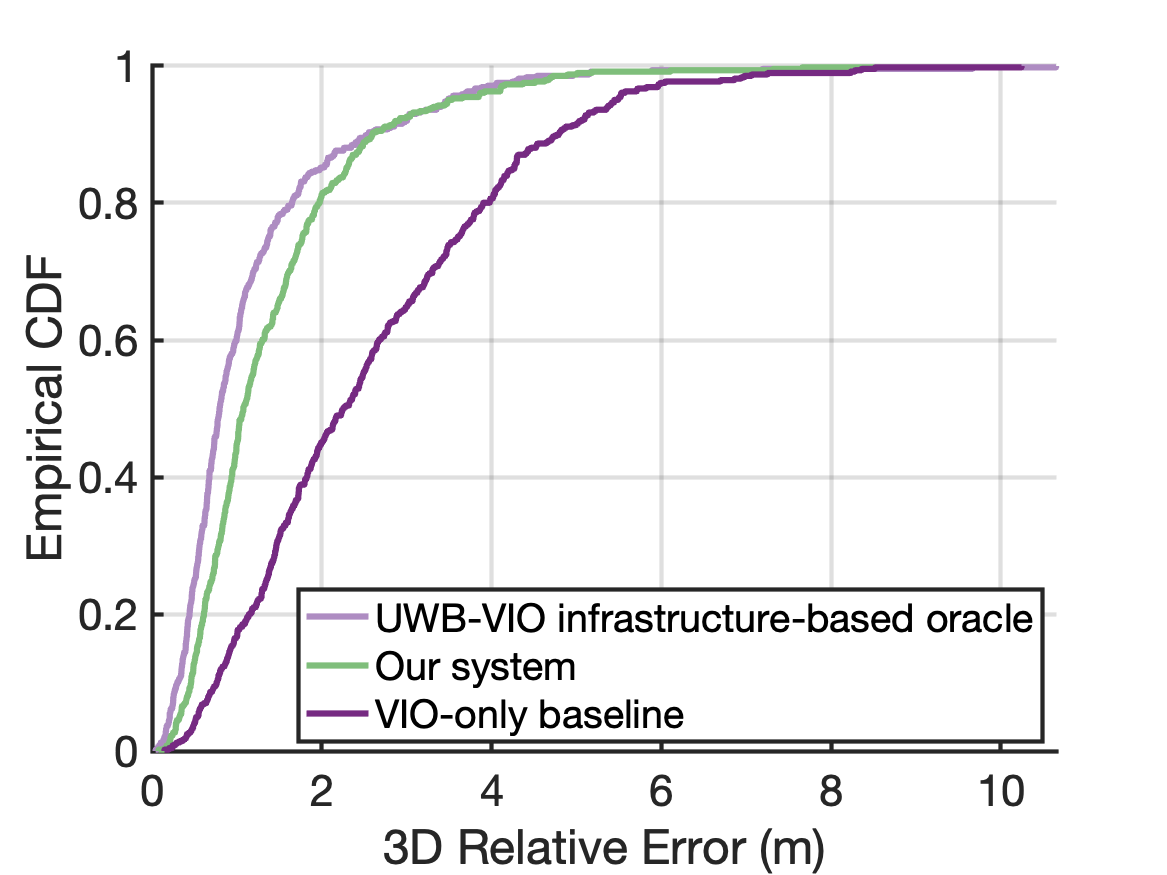}
        \caption{3D Relative Location Error}
       \label{fig:mainLocErr}
\end{minipage}
\begin{minipage}{0.49\linewidth}
        \centering
        \includegraphics[width=\linewidth]{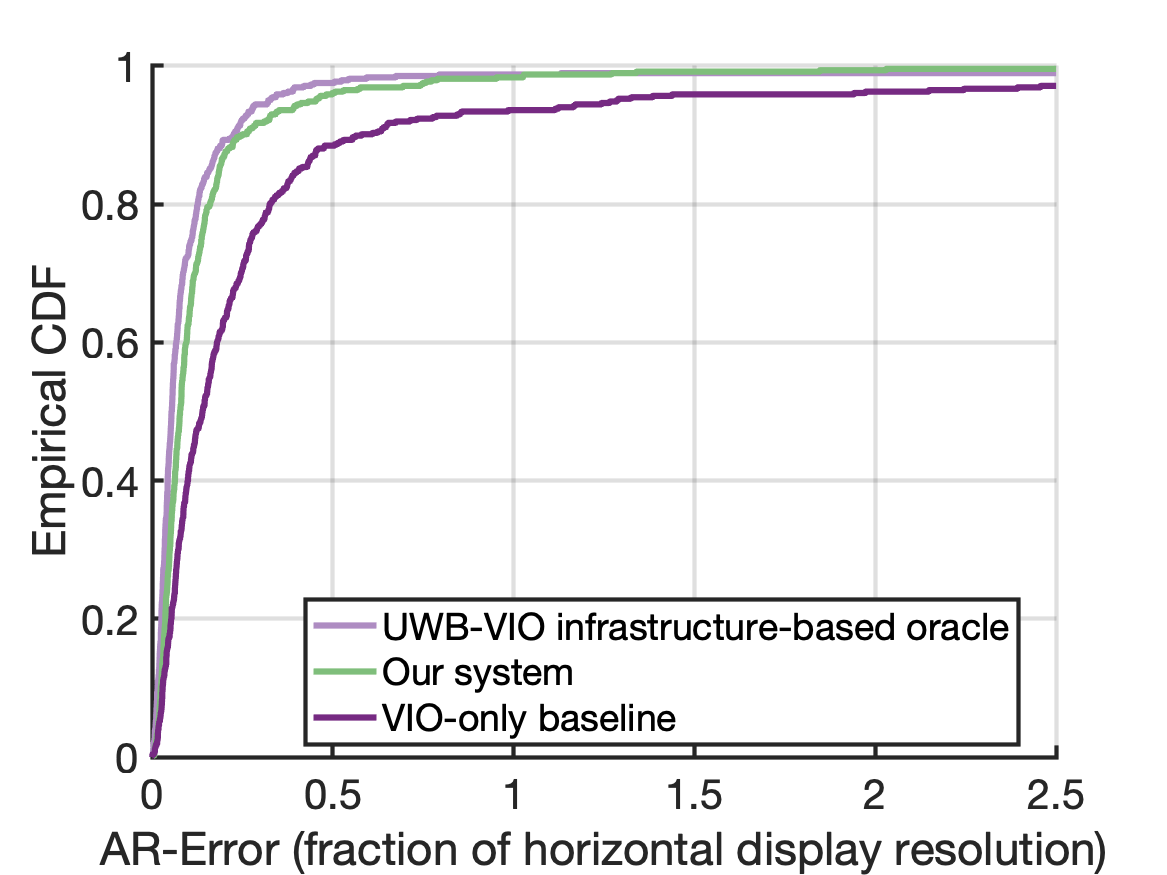}
        \caption{AR-specific Display-Proportional Error Metric. }
            \label{fig:mainARErr}
\end{minipage}
   \end{figure}%
   
\subsection{Localization Accuracy} \label{sec:loc_acc}
We evaluate the localization accuracy of \sysname across our evaluation dataset with 5 mobile users, on both single and multiple floors, and with a mixture of LOS and NLOS situations. Figure \ref{fig:mainLocErr} shows the overall 3D relative localization error and compares it with our two baseline approaches. \sysname achieves a median 3D error of 0.9 m, compared to 2.5 m and 0.8 m in the VIO-Only baseline and UWB-VIO oracle, respectively. We can see that \sysname outperforms VIO-Only baseline by leveraging the UWB ranging and collaborative pose estimation which eliminates drift over time. In addition, \sysname achieves relatively similar accuracy to the UWB-VIO oracle, which relies on pre-installed infrastructure and {\em a priori} knowledge of beacons for trilateration that is unnecessary for \sysname.



As mentioned in Section \ref{sec:metric}, the 3D geometric error does not necessarily quantify the localization performance specific to AR application, hence our proposed AR metric, display-proportional error, which captures the relative object displacement error on the screen. Figure \ref{fig:mainARErr} compares the AR performance of the three methods using our defined metric, and shows that \sysname can satisfy a high AR quality in 99\% of cases with less than 0.5 fractional error on display and a median quality 0.1 fractional error. This means that the users are able to steer in the right direction toward a user 99\% of the time across varying ranges and angles.

\subsection{Error vs. Separation Distance}
\begin{figure}[t]
    \centering
     \begin{minipage}{0.49\linewidth}
        \includegraphics[width=\linewidth]{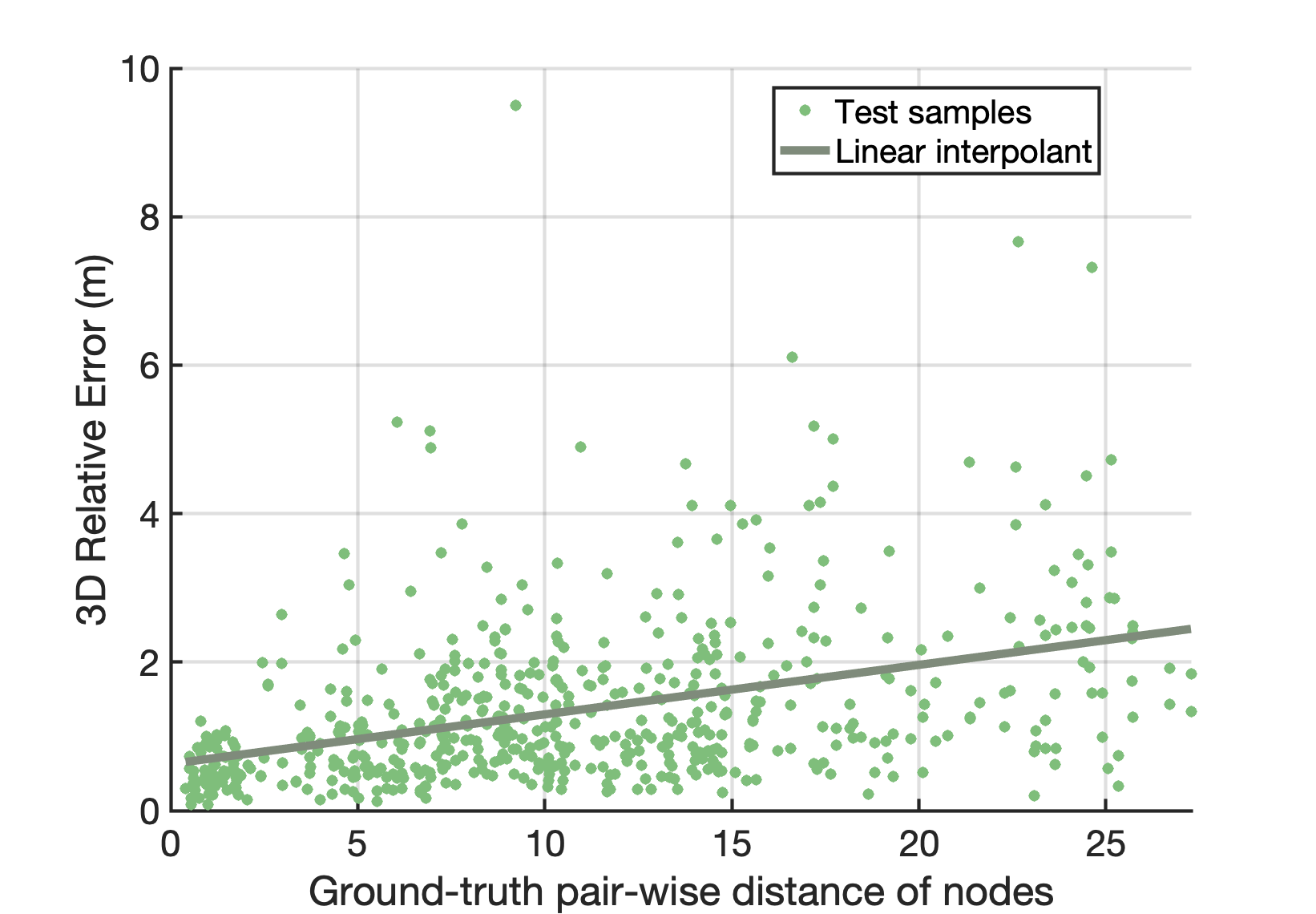}
        \caption{3D error vs. Distance}
        \label{fig:distLoc}
    \end{minipage}
    \begin{minipage}{0.49\linewidth}
           \centering
        \includegraphics[width=\linewidth]{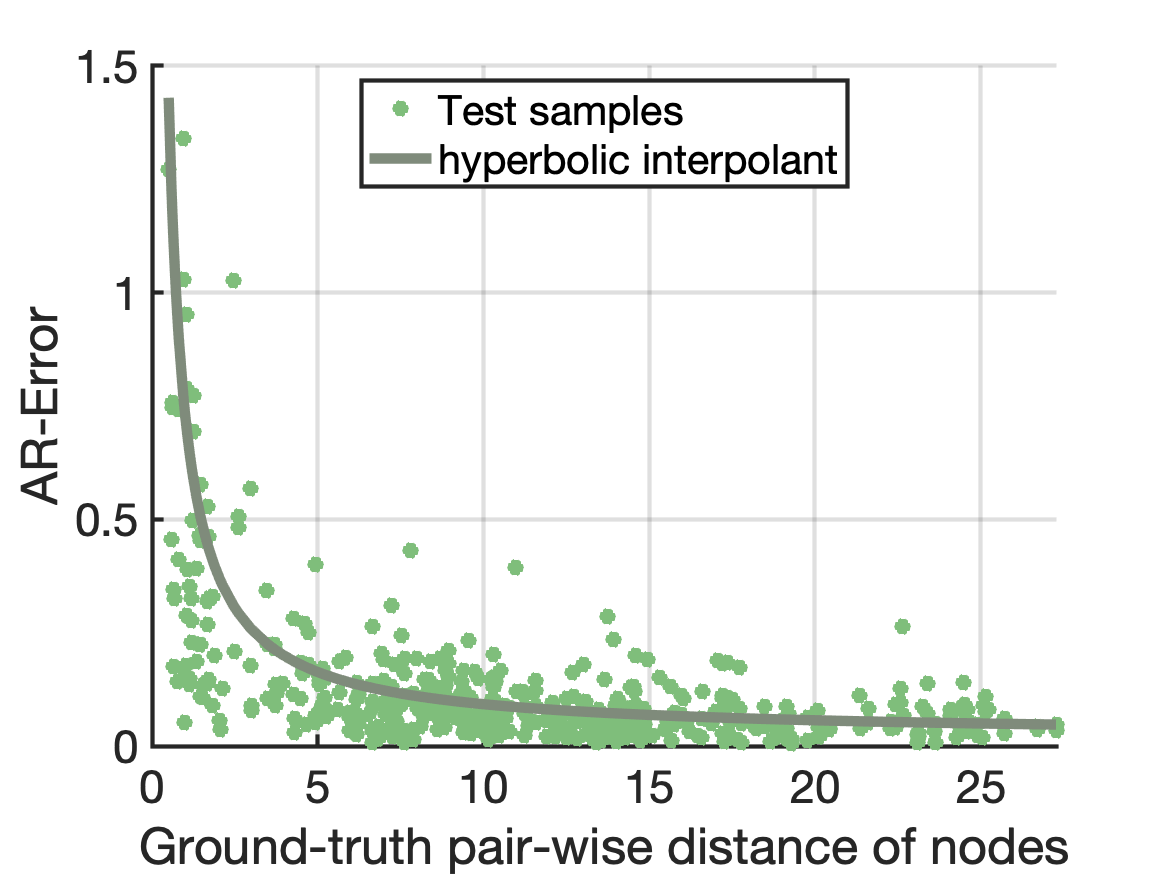}
        \caption{AR-error vs. Distance}
           \label{fig:distAR}
    \end{minipage}
\end{figure}

Next, we evaluate \sysname's performance as a function of distance. The ground-truth relative distance of users varies between 0.2 m to 27 m including many instances of completely NLOS. Figure \ref{fig:distLoc}-a demonstrates the 3D relative error of each sample test (any pair of users at every 5 s interval) grouped by the ground-truth pair-wise distances. As we can see, error in positioning tends to increase slightly with distance, either due to UWB nodes going out of range or inherent VIO drift. However, unlike geometric error, DPE actually improves with distance. This suggests that a visual display showing an overlay with faraway users' locations would still be effective at portraying those users' locations.  

It can be seen in Figure \ref{fig:distLoc} that there is a linear correlation between error and true distance of nodes and that error stays below 10\% the distance on average. This means that at 20 m, the average error is less than 2 m, which is a reasonable amount of error when localizing 2 users in a building with NLOS. This can be better captured with the AR-metric shown in Figure \ref{fig:distAR} with a decreasing error over extended ranges. As we can see, the trend of geometric error and AR error with respect to true distance is opposite, confirming the AR-specific behavior of localization systems explained in Section \ref{sec:metric}.


\subsection{Drift Over Time}

\begin{figure}
    \centering
    \begin{minipage}{0.49\linewidth}
        \includegraphics[width=\linewidth]{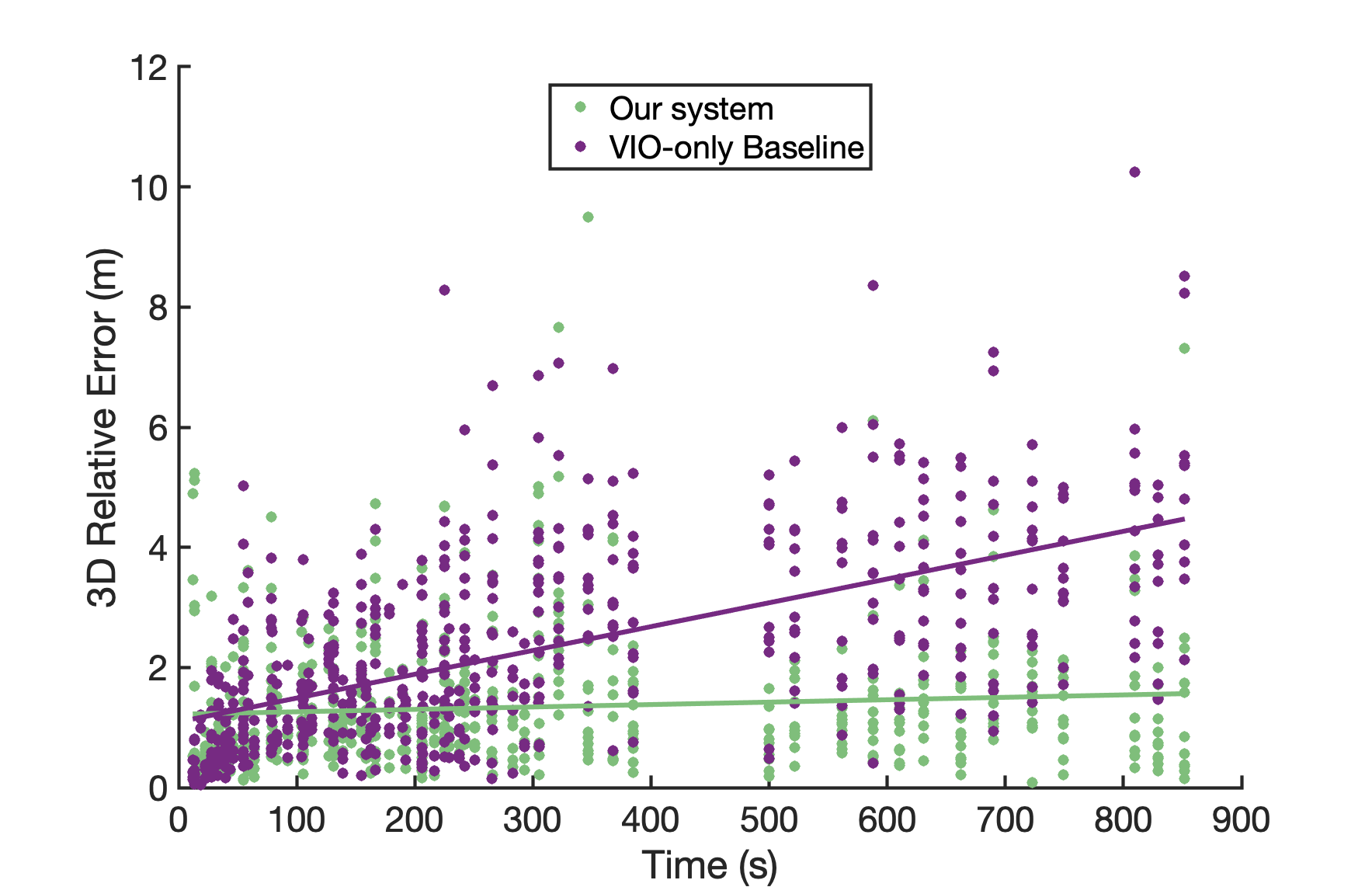}
        \captionsetup{width=.9\linewidth}
        \caption{3D error over time. Note how VIO drifts leading to increasing errors, while \sysname preserves a uniform accuracy by leveraging UWB ranging between users.}
        \label{fig:overTime}
    \end{minipage}
    \begin{minipage}{0.49\linewidth}
        \centering
        \captionsetup{width=.9\linewidth}
        \includegraphics[width=\linewidth]{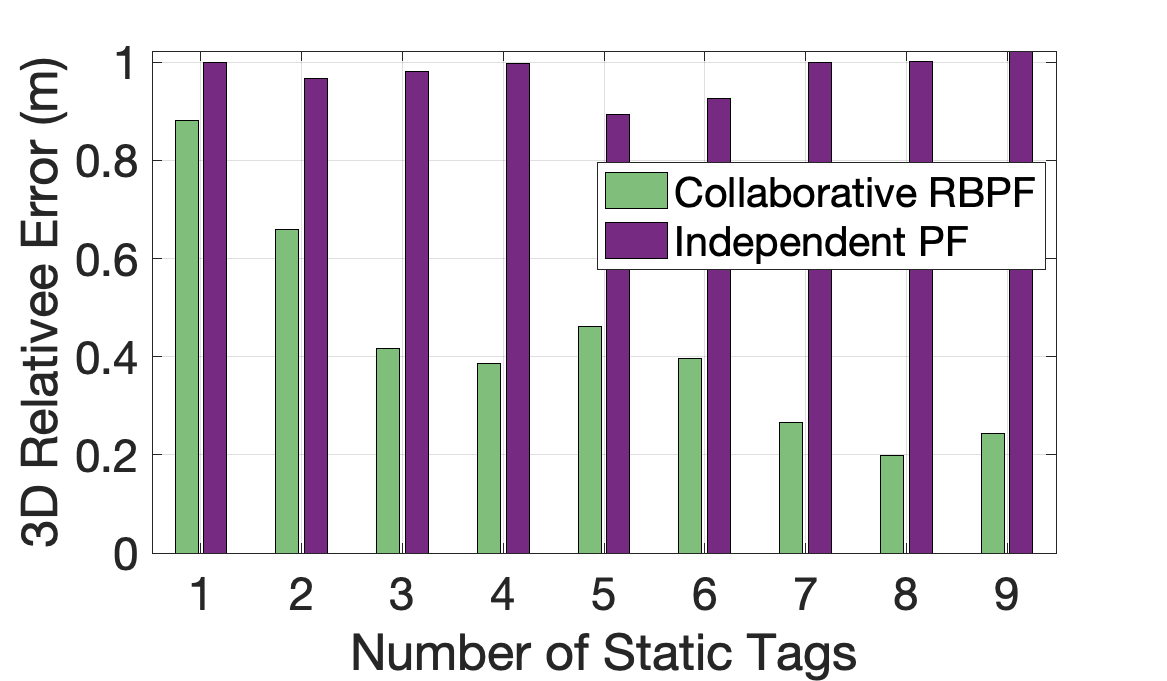}
    
        \caption{Impact of collaborative Localization.  Using a collaborative approach improves the localization performance as the number of nodes increases.}
        \label{fig:collabImpact}
    \end{minipage}
\end{figure}

In many localization systems, including VIO tracking, error increases with time. Dead reckoning systems have inherent drift that is inevitable, and small errors in individual state estimation accumulate over time. As seen in Figure \ref{fig:overTime}, \sysname does not have this problem, and keeps an almost constant error distribution throughout every experimental run, while VIO has linear drift over time. This success can be attributed to the use of UWB ranging between nodes to keep the drift bounded to within the UWB error.






\subsection{Impact of Collaborative Localization}
\label{sec:collab_impact}
To evaluate the impact of our collaborative particle filter formulation, we compare the 3D relative error of the naive independent PF and collaborative RBPF, explained in Section \ref{sec:PF}. To isolate the impact of other parameters, including user mobility, number of users, etc, we perform controlled experiments with a single user and 9 static tags deployed for baseline comparisons. Then we estimate the relative location of static tags with respect to the user for different subsets of tags changing from 1 to 9 randomly selected tags. As seen in Figure \ref{fig:collabImpact}, collaborative RBPF has a clear advantage over Independent PF. While with 1 static node, they are very similar in 3D relative error of around 0.9 m, independent PF localization remains at this error while the error of collaborative RBPF decreases from 0.85 m to 0.22 m with the addition of static tags.

This was expected as collaborative RBPF takes advantage of other system nodes' estimates. With this method, the drift of the mobile node is able to be somewhat mitigated by the averaging of noise across measurements to multiple other nodes. As more nodes are able to perform these "averaging corrections" together, the localization system is able to converge to a more precise estimate than it could with nodes localizing individually. As a side conclusion, we can leverage this feature to further improve the localization performance by deploying some static UWB tags with unknown locations. For example, in a first response operation, the users can deploy some static nodes at random locations as they move around the building to enhance their relative localization performance. Even though \sysname can operate completely infrastructure-free, it can nicely integrate with the infrastructure if one is present. 

\section{Sensitivity Analysis} \label{sec:sensitivity}
\makeblue{In this section, we elaborate on the computational overhead of \sysname's collaborative localization algorithms. We also describe additional tests we performed in other campus environments and evaluate the sensitivity of \sysname to varying user mobility patterns and NLOS conditions in these environments.}

\subsection{Computational Overhead} \label{sec:comp_overhead}
Real-time convergence and computation are two of the critical requirements of an AR localization system specially in mobile applications. Compared to independent particle filtering, our collaborative formulation achieves higher accuracy at the cost of higher computational overhead. Table \ref{tab:compute}, however, shows that \sysname can still operate on a wide variety of platforms and in practice, converges in real-time. It should be noted that our implementation is not heavily optimized, and our compute time includes significant system overhead. The key takeaway is that the run-time overhead increases almost linearly with the number of users.

\begin{table}[b]
	\centering
 \begin{tabular}{|c| c| c| c|c|} 
 \hline
 Number of users & 2 & 3 & 4 & 5  \\ 
  \hline
 CPU Usage (ms) & 2.3\% & 8.0\% & 16\% & 25\%\\ 
 \hline
 Memory Usage (MB)  & 4 & 8 &  12 & 16 \\
 
\hline

\end{tabular}
\caption{Single threaded runtime performance on 2.4GHz i7 CPU} 
\label{tab:compute} 
\end{table}




\makeblue{
\subsection{Performance Across Diverse Environments} \label{sec:environments}
In addition to the multi-story building tests described in Section \ref{sec:evaluation}, we also performed a series of tests across several other environments under different conditions, many of which are shown in Figure \ref{fig:moreEnv}. These environments included:
\begin{itemize}
    \item A "busy" office (with furniture being moved and lights being turned on and off to simulate ordinary office commotion)
    \item A campus cafe with a large atrium and spiral staircase
    \item A hallway intersection near some elevators inside a brick building
    \item A dimly lit parking garage with height variation and lots of concrete and metal blocking line of sight.
    \item An outdoor area between campus buildings
\end{itemize}
The results of these experiments are shown in Figure \ref{fig:sensitivity}. We see that performance is consistent across all of these environments, with the parking garage performance suffering slightly due to the heavy NLOS conditions and low light. Note that the performance in all of these environments is slightly better than the primary multi-story building test, which was the most challenging due to its immense scale.
}

\subsection{Impact of Mobility Pattern} \label{sec:mobility_pattern}
Another factor affecting the performance of \sysname's localization accuracy is the high dynamics of the environment and mobility of users. To this end, we compared the system performance in 3 different walking scenarios: (1) when users were walking in pairs, which represents the near-best performance as the collaborative algorithm can take advantage of clean ranging estimates between each pair of users walking near each other, (2) normal walking when users randomly move in the space with a usual walking speed, (3) when all the users were performing fast movements such as running, jumping, crawling, etc, for the purpose of stress testing the algorithm. Figure \ref{fig:sensitivity}-b confirms the expected trend for different walking scenarios, and demonstrates that \sysname is resilient to fast motions and therefore suitable for applications that involves fast motions, such as rescue operations or gaming.

\subsection{NLOS Performance} \label{NLOS performance}
\begin{figure}
    \centering
    \includegraphics[height=2.1cm]{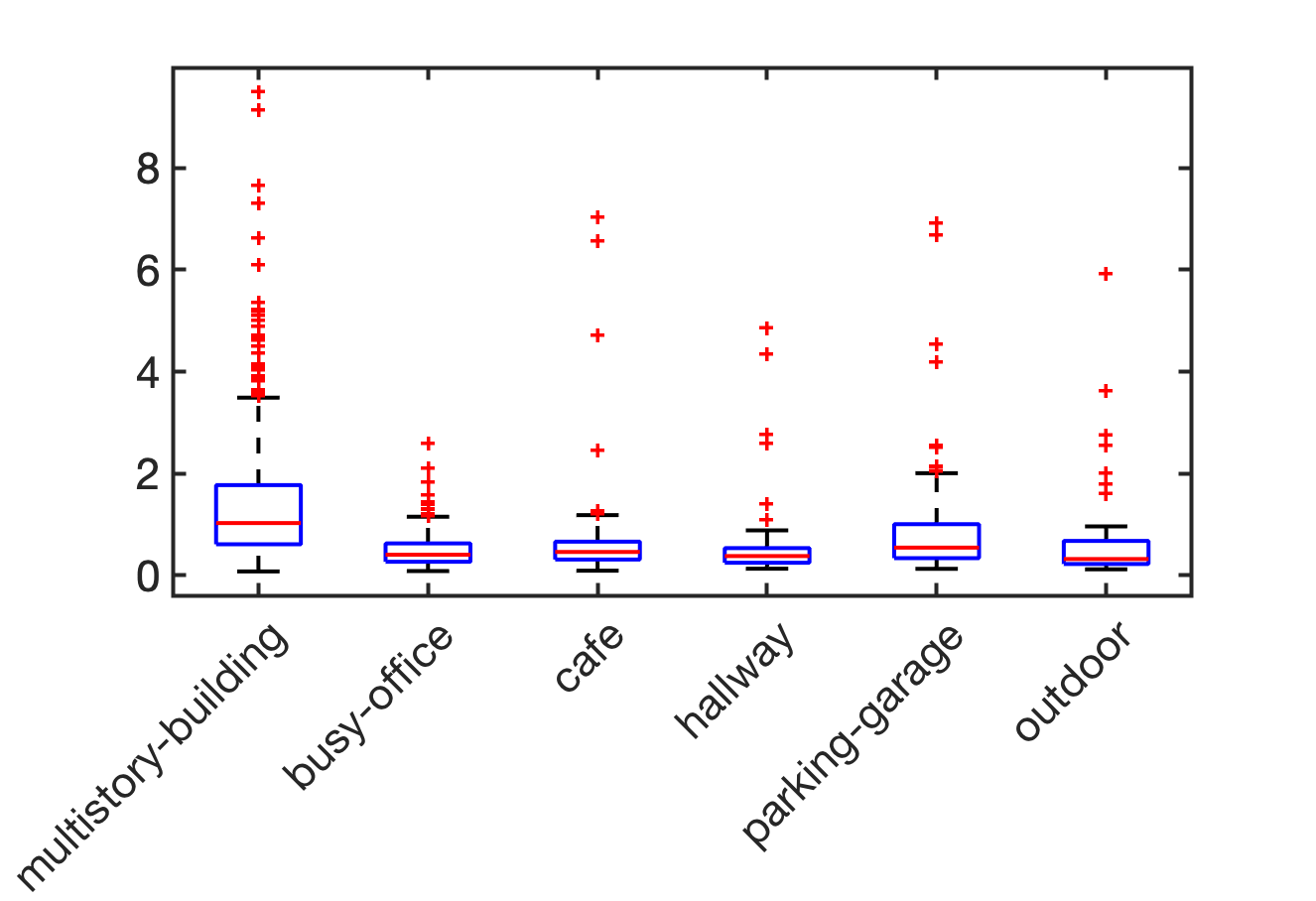}
    \includegraphics[height=2.1cm]{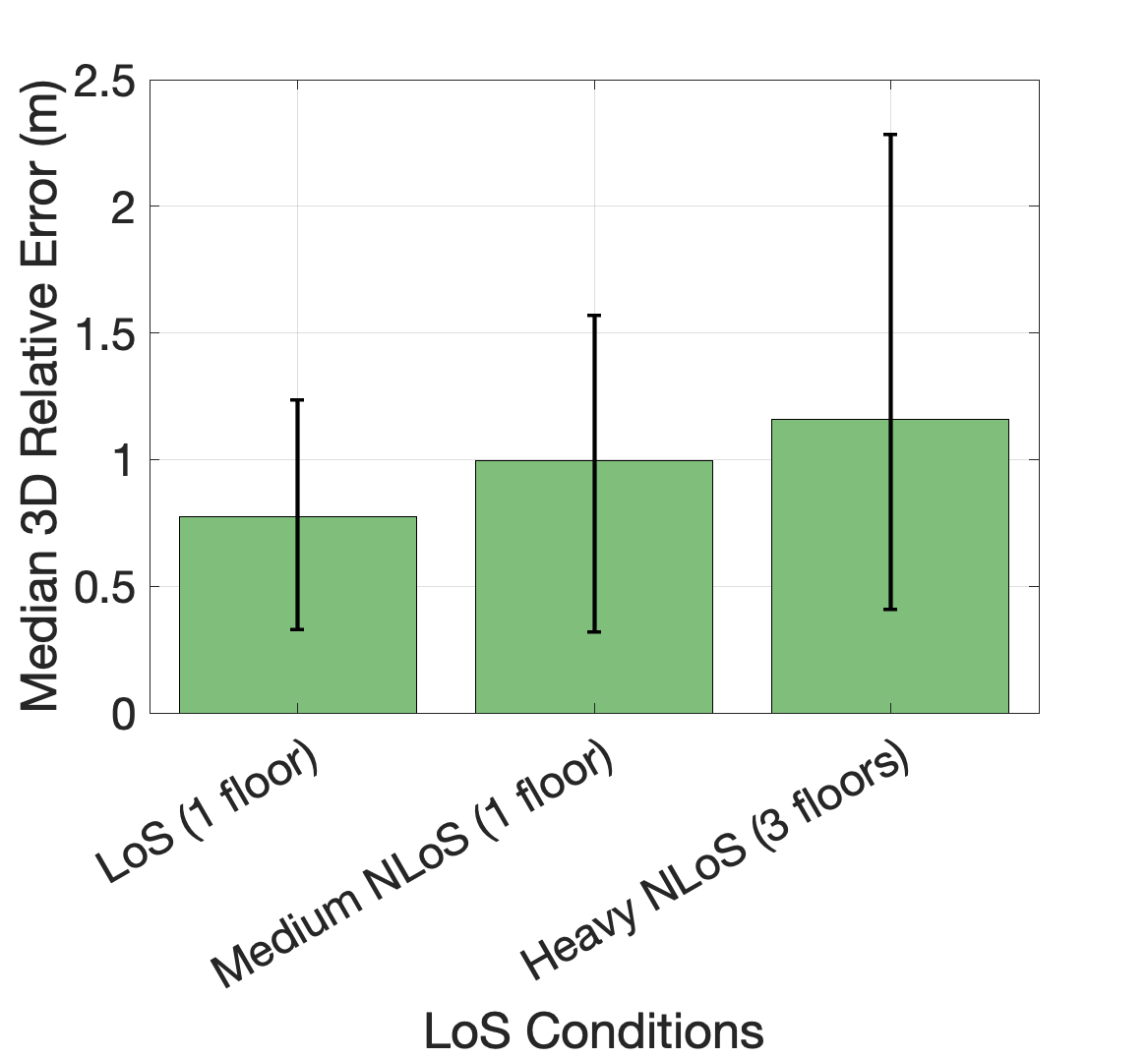}
    \includegraphics[height=2.1cm]{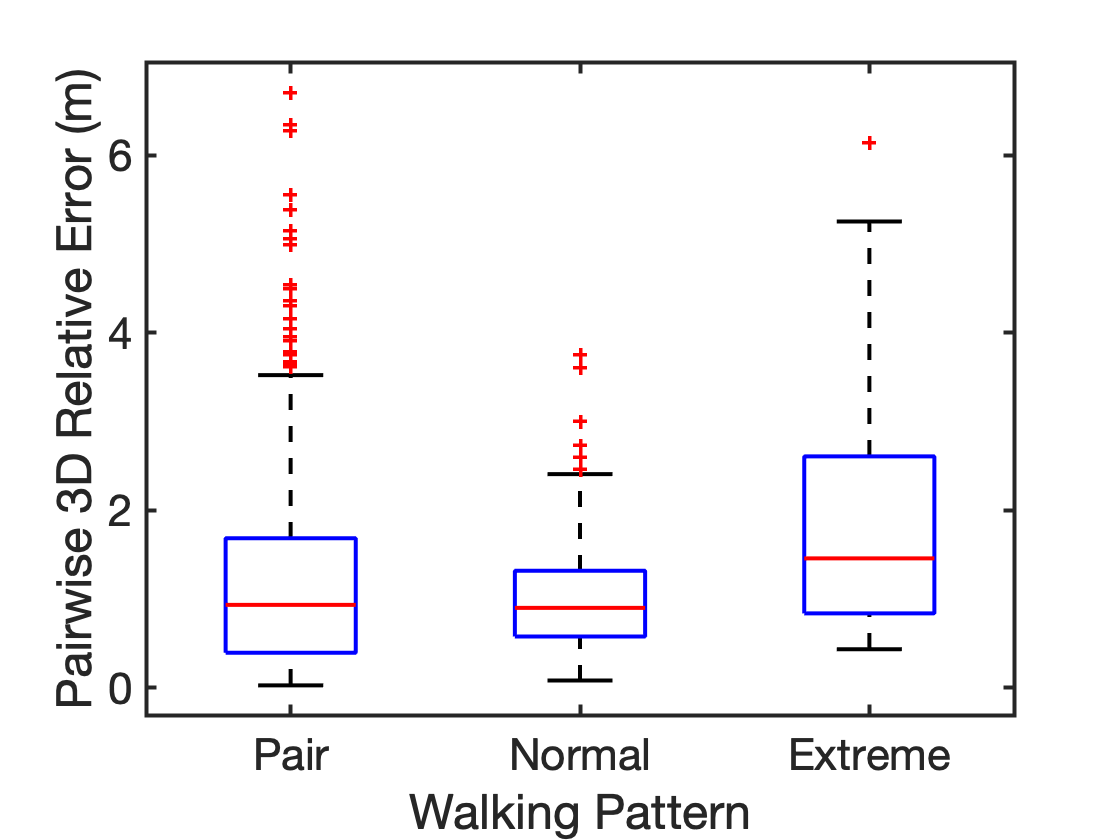}
    \caption{\sysname preserves high accuracy in different lighting, mobility, or NLOS conditions as well as walking patterns.}
    \label{fig:sensitivity}
\end{figure}

Next, we study \sysname's localization performance in NLOS scenarios. Previous analyses shows that UWB ranging degrades in complete NLOS \cite{rajagopal2019improving} due to noisy Time of Flight (TOF) estimates that mainly capture multipath reflections instead of the direct distance between nodes. To evaluate this effect, we performed 3 different experiments with different levels of NLOS scenarios. The first experiment includes 5 users that walk mostly in LOS of each other, all on the same floor. We then repeated this experiment, while users were walking in a larger space including both LOS and NLOS conditions. Finally, we performed the experiment while users were spread out across 3 floors with some heavy NLOS conditions such as multiple concrete walls between users, or being apart by more than 1 floor. As we can see in Figure \ref{fig:sensitivity}-b, the 3D relative localization drops slightly with the increase of NLOS conditions, but we can still maintain a median accuracy of ~1 m even in NLOS and extended ranges over 10-20 m. 

\section{Discussion}

\makeblue{In this section, we discuss the mechanisms to relax assumptions made in our current implementation of \sysname and the potential future extensions. }

\vspace*{0.05in}\noindent \makeblue{\textbf{Number of users}: While the current evaluations are done with maximum of 5 mobile users and 9 stationary nodes, \sysname's collaborative approach can significantly benefit from larger number of users to improve the localization accuracy (as shown in Figure \ref{fig:collabImpact}). This is mainly due to drift mitigation by averaging the noises across measurements to multiple nodes. However, the higher performance will be achieved at the expense of higher computation and communication overhead. While our current implementation is not optimized for scalability, this trade-off can be balanced by using heuristics that clusters users based on their proximity. Designing a scalable and real-time version of \sysname is part of our future work.  }

\makeblue{On a related note, our current algorithm only uses the direct measurements from each neighbor without sharing any higher-level state information (such as estimates of other users etc). This reduces the communication complexity and computational overhead at the expense of slightly lower performance. Depending on the application and required localization performance, the collaboration capabilities of \sysname can be further enhanced by sharing ranging and pose estimations between users.}

\vspace*{0.05in}\noindent \makeblue{\textbf{User Interactions}: In practice, \sysname works best when users occasionally pass near each other, resulting in high-confident ranging and particle filter updating. So, the algorithm cannot benefit from collaboration if users are at the limits of the UWB range (100m in LoS and about 30m in sever NLoS). To avoid the performance degradation, one can add (arbitrarily placed). Such "breadcrumb dropping" techniques \cite{li2018automatic} has been widely used in rescue operations or first response operations, and are compatible with \sysname. It is worth noting that without close interaction the system will converge, just over a longer period of time or at a lower accuracy. }

\vspace*{0.05in}\noindent \makeblue{\textbf{Camera Occlusion:} A limitation that is common among vision-based IMU and localization methods is sensitivity to low visibility conditions, such as smoke-filled rooms or extreme darkness, as would be commonplace for firefighters. Our current experiments show that \sysname is resilient to partial camera blockages by leveraging the UWB ranging between users. An ultimate solution could be to replace vision with RF imaging technologies, such as millimeter wave (mmWave) \cite{mmWaveEKF} or Infrared cameras. As part of our future work, we are integrating our infrastructure-free relative localization with mmWave cameras that are not affected by smoke.  } 

\vspace*{0.05in}\noindent \makeblue{\textbf{Groundtruth Collection:} One of the main challenges of any localization research is groundtruth collection especially in mobile indoor setups where GPS does not work well. As such, we designed a novel mechanism using April tags and coordinated synchronous check pointing that guided users as they walked through our test environments. However, When collecting the ground truth, participants had to wait until all of them had high-confidence ground truth measurements for synchronization. While we don't expect this approach to impact the evaluation, it could slightly improve the UWB ranging as the users will be static for a few seconds when collecting GT. However, given the high sampling rate of UWB ranging and the heavily NLoS conditions of our experiments, we imagine this impact to be negligible. 
}




\section{Conclusion}

This paper proposes \sysname, a collaborative AR localization system that allows multiple users to estimate the relative 6DOF position of each other in real-time. This system is free of infrastructure; is robust to environment dynamics and NLoS conditions; and maintains relatively low computational complexity to reduce power and update time. 
\sysname uses a variant of RBPF to perform state estimation of the nodes jointly by using UWB ranging and VIO tracking. \sysname then displays the tracked nodes in AR on the tablet in the coordinate frame of the user. Using the AR application, users can see where others are in the building despite walls, floors, and other obstacles creating NLOS. We also present an AR metric that captures the quality of localization with respect to the user's display specifications, and is optimized for augmented reality applications. 

As future work, we are interested in using the \sysname approach to bootstrap and correct mapped locations within fixed infrastructure systems.  There is the potential to create a hybrid infrastructure-based and infrastructure-free AR localization environment that could provide the best of both worlds where rapidly deployed relative content could persist in the environment once fixed infrastructure is encountered.

\bibliographystyle{ACM-Reference-Format}
\bibliography{References}

\end{document}